\documentclass[a4paper,11pt]{report}
\newcommand{\linespacing}{1.5}
\newcommand{\al}{\alpha}
\newcommand{\Om}{\Omega}
\newcommand{\tl}{\tilde}
\newcommand{\beq}{\begin{equation}}
\newcommand{\eeq}{\end{equation}}

\newcommand{\om}{\omega}
\newcommand{\gm}{\gamma}
\newcommand{\dl}{\delta}
\newcommand{\dau}{\partial}
\newcommand{\p}{^\prime}
\newcommand{\ep}{\epsilon}
\newcommand{\pl}{\parallel}
\newcommand{\pp}{\perp}
\newcommand{\tb}{\textbf}
\renewcommand{\baselinestretch}{\linespacing}
\usepackage{natbib}
\usepackage{graphicx,color}
\usepackage{epstopdf}
\usepackage[british]{babel}
\usepackage{amsmath}

\usepackage[a4paper,top=2.5cm,bottom=2.5cm,left=4cm,right=2cm,headsep=10pt]{geometry}
\usepackage[colorlinks,pagebackref,pdfusetitle,urlcolor=blue,citecolor=blue,linkcolor=blue,bookmarksnumbered,plainpages=false]{hyperref}

\begin{document}
\pagenumbering{roman}
\thispagestyle{empty}

\begin{flushright}
\includegraphics[width=6cm]{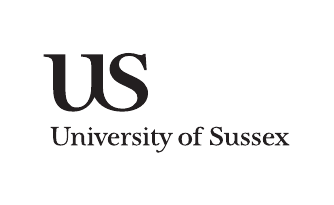}
\end{flushright}	

\vskip40mm
\begin{center}
\huge\textbf{Production of Gravitational Waves in the Early Universe}
\vskip2mm
\LARGE\textit{From turbulence triggered by first-order phase transitions}
\vskip5mm
\Large\textbf{Yashmitha Kumaran}
\normalsize
\end{center}

\vfill
\begin{flushleft}
\large
Submitted for the degree of Master of Science \\
University of Sussex	\\
16th August 2018
\end{flushleft}		

\chapter*{Summary}
This project is aimed at studying the first-order phase transitions, that is presumed to have ensued in the early universe, and its consequences on the primordial gravitational waves. The effects of bubble nucleation, growth and coalescence are reviewed. The resulting first-order phase transition is taken as the source of the gravitational waves that were produced, in order to determine the energy density, amplitude and frequency spectra of the relic gravitational wave background. This is accomplished by modelling the first-order phase transition as a turbulent fluid and employing relativistic hydrodynamic equations to estimate the required physical quantities.
 
Two models are majorly studied for all the analysis done in this project. Both models compute the necessary gravitational wave spectra using the exponential Kraichnan function as the temporal decorrelation function. Also, both models contemplate the turbulence in the flow of plasma to be stationary, obeying the conditions dictated by the Kolmogorov turbulence. However, the first model uses a de-coherence function that depends on the wavenumber and time, while, the second model uses the top hat correlation, to compute the anisotropic stress. The new model introduced here adheres to the freely decaying turbulence model, but employs the time dependant de-coherence function in its computations. The two models are reproduced and their essential functions are mimicked to obtain anisotropic stress from the velocity power spectrum for unequal times, and hence, the density, amplitude and frequency spectra. The long term goal of this project is to increase the possibility of the gravitational wave detection in the future.

\chapter*{Preface}
Chapter 3 begins with focusing on the work done for the calculation of the gravitational wave power spectrum with the help of longitudinal waves. I, in conjecture with my supervisor, have modified the same to transverse waves, so as to account for the gravitational waves more precisely. It further concentrates on the computation of the amplitude and frequency spectrum for the stationary turbulence model of the first-order phase transitions using a de-coherent function for the velocity power spectrum. Chapter 4 emphasizes on the power spectrum of the gravitational wave energy density for the freely decaying turbulence model with the top hat approximation for the correlation of the velocity power spectrum. In chapter 5, I have reproduced the results of the two models given by the respective researchers and applied them to tailor the new model, wherever applicable.
 
This analysis is an intersection of these models in an attempt to account for the assumptions and approximations for some parameters like the source duration, de-coherence function, etc. The computation done for this new model is exclusively our own effort.

\section*{Author's note:}
This dissertation was originally written and submitted in 2018 to satisfy the requirements for the Masters degree. The work was primarily intended as a pedagogical study that reviews the production of gravitational waves in the early universe from first-order phase transitions, with the aim of outlining the relevant literature and theories. Following the completion of this dissertation, my supervisor and collaborators published research articles that further developed this framework comprehensively; refer: \cite{Hindmarsh:2019phv}, \cite{Cutting:2020nla}, and \cite{Hindmarsh:2020hop}. This is a slightly revised edition of the original submission, prepared for public archiving and documentation purposes. Scientific context and conclusions remain the same; revisions are limited to improvements in presentation \& clarity and formatting added to this note. The computational scripts found in the appendix of the master version are not included in this public version.

\newpage
\pdfbookmark[0]{Contents}{contents_bookmark}
\tableofcontents
\listoffigures
\phantomsection
\addcontentsline{toc}{chapter}{List of Figures}

\newpage
\pagenumbering{arabic}

\chapter{Introduction}
\label{chap:1}
Ever since Oliver Heaviside discovered the analogy given by the inverse-square law in electricity and in gravitation, the existence of gravitational waves has been an intriguing question. Henri Poincare broadened this analogy further by relating accelerating charges and accelerating masses. In the meantime, Albert Einstein had comprehended the ‘ripples in the metric of spacetime’ in his “General Theory of Relativity” in the year 1916. According to his theory, accelerating massive objects in space produce waves that disrupts spacetime, travels at the speed of light away from its origin as ripples and carries energy and information about its source in the form of gravitational radiation. While, In Einstein’s words, Poincare was a pioneer in relativity, and published a shorter version of a paper on gravitational waves just three months before, and a longer version soon after, the publication of the General Theory of Relativity, the excellence of the work from Einstein delayed Poincare being widely recognised for his work on Lorentz transformations until Hendrik Lorentz acknowledged him in one of his publications in 1921.
 
Although Einstein’s mathematics contrasted the Newtonian theory of gravitation that the speed of the propagation of gravity is infinity, it was detected for the first time in the year 1974 from a binary system of a neutron star and a pulsar called the Hulse-Taylor binary. A hundred years after Einstein’s prediction, the existence of such waves was detected in September, 2015, from two merging black holes that was a billion light years away from earth, and proved in February, 2016, by the researchers at the Laser Interferometer Gravitational-wave Observatory. This first detection at LIGO, known as GW150914, confirmed Einstein’s theory and expanded the scope of studying the universe and its history. Over the decades, the concept of gravitational waves has taken various forms and has been discovered to be arising from colliding black holes, merging neutron stars, supernova explosions, etc. But, the gravitational waves that were possibly produced in the early universe during the phase transitions is the subject that this project is focused on.
 
The universe has immensely evolved to get to this point of how it is seen today. In these billions of years, the role of matter has been – and still being – a major factor of building, influence and growth.  When it is traced back to the era of violent processes in the infant universe, the extreme conditions (especially temperature and density) are found to have had a huge impact on the symmetry of matter in the form of phase transitions, that eventually resulted in its present day properties. At the moment of creation, matter is believed to have been in unified phase, which was then subjected to a series of phase transitions driven by the decrease in temperature until several of its components decoupled. If the phase transitions during the adiabatic expansion is of first order, phase bubbles are thought to have formed. Anisotropic stresses would have been created due to the nucleation and collision, serving as a potential source \citep{Caprini:2015tfa} for the production of gravitational waves.
 
There could be other possible sources in the later time frames of the expanding universe. However, symmetry getting restored at extremely high temperatures implies the significance of symmetry breaking, suggesting that the phenomenon of phase transition is comparatively more plausible to have occurred as a source of gravitational waves. Second order phase transitions are not a choice either because of their continuous nature.\footnote{Referred from \url{http://www.damtp.cam.ac.uk/research/gr/public/cs_phase.html}}
While both these transitions are assumed to have ultimately led to inflation, the dramatic first order phase transitions are chosen over the smooth second order phase transitions. The Higgs boson particle discovered at the LHC proved the breaking of symmetry, a symmetry that was driven by a scalar field, enhancing the odds of the first order phase transitions.
 
In first order transitions, bubbles of a new phase form in the midst of the old phase. These vacuum bubbles grow, expand and collide with each other until the old phase disappears. As the phase transition culminates, the colliding bubbles convert the whole universe into a broken phase. The collisions also induce an anisotropic stress due to the loss of spherical symmetry of the bubble wall, which is non-zero, and instigate compression waves and turbulence. The transition is said to be complete when the bubble merge and the old phase completely transforms into the new phase. The main factor involved in this process of symmetry breaking is temperature.
 
Calling condensed matter physics for an explanation, the force carrier gauge bosons are massive at low energies breaking the symmetry, and massless at high energies restoring the symmetry. The potential of the scalar field driving the symmetry is seen to vary as mass-squared, provided that the scalar field takes a constant value determined by its interactions. Here, the sign of mass-squared term is crucial: when it is positive, the potential reaches a minimum – called vacuum – at the origin, that corresponds to the energy minimum of the system at constant fields, restoring the symmetry. On the other hand, when the mass-squared term is negative, the potential has another non-zero minimum along with its minimum at the origin, retaining the mass term for the gauge bosons, and hence, breaking the symmetry.
 
In case of phase transitions, the shape of the potential is required to be changed through temperature corrections for the scalar field to gain a positive contribution. This is done by defining a critical temperature. Applying the conclusions from condensed matter physics, if the temperature term is greater than the critical temperature, symmetry exists; else, the symmetry is broken. When the temperature cools down to below the critical temperature rapidly (or quenches) from the highest free-energy phase at high temperature, then the system is trapped in the high free-energy phase, becoming metastable.
 
Quenching to the lowest free-energy phase is a discontinuous process that involves bubbles from the low free-energy phase nucleating in the metastable phase. The discontinuity clearly suggests that this is a first order phase transition. The consequence of these coherent, nonlinear fluctuations being gravitational waves is the fundamental idea behind this project, with the final objective to determine the amplitude and frequency spectrum of the resulting gravitational waves.

\newpage
\chapter{Literature survey}
\label{chap:lits}

\section{Phase transitions: mathematical representation}
The concept of phase transitions plays a major role in solving the cosmological mysteries. One such example is the resolution of the smoothness problem of the Big Bang model by the introduction of inflationary cosmology. Building on that, gravitational waves can be seen as the signature of the early universe during the phase transitions, if and when they had occurred. This highlights the importance of symmetry breaking for a scalar field, $\phi$ and a potential $V(\phi)$, with mass, $m$. Mathematically, the potential of the symmetry-driven scalar field \citep{Gleiser:1998kk} can be expressed as:

\beq
V \left(|\phi|^2 \right) = m^2 |\phi|^2 + \lambda_{\phi} |\phi|^4
\eeq
where, the scalar quantity $\lambda > 0$ is a self coupling constant.
 
\begin{figure}
\centering
\includegraphics[width=10cm]{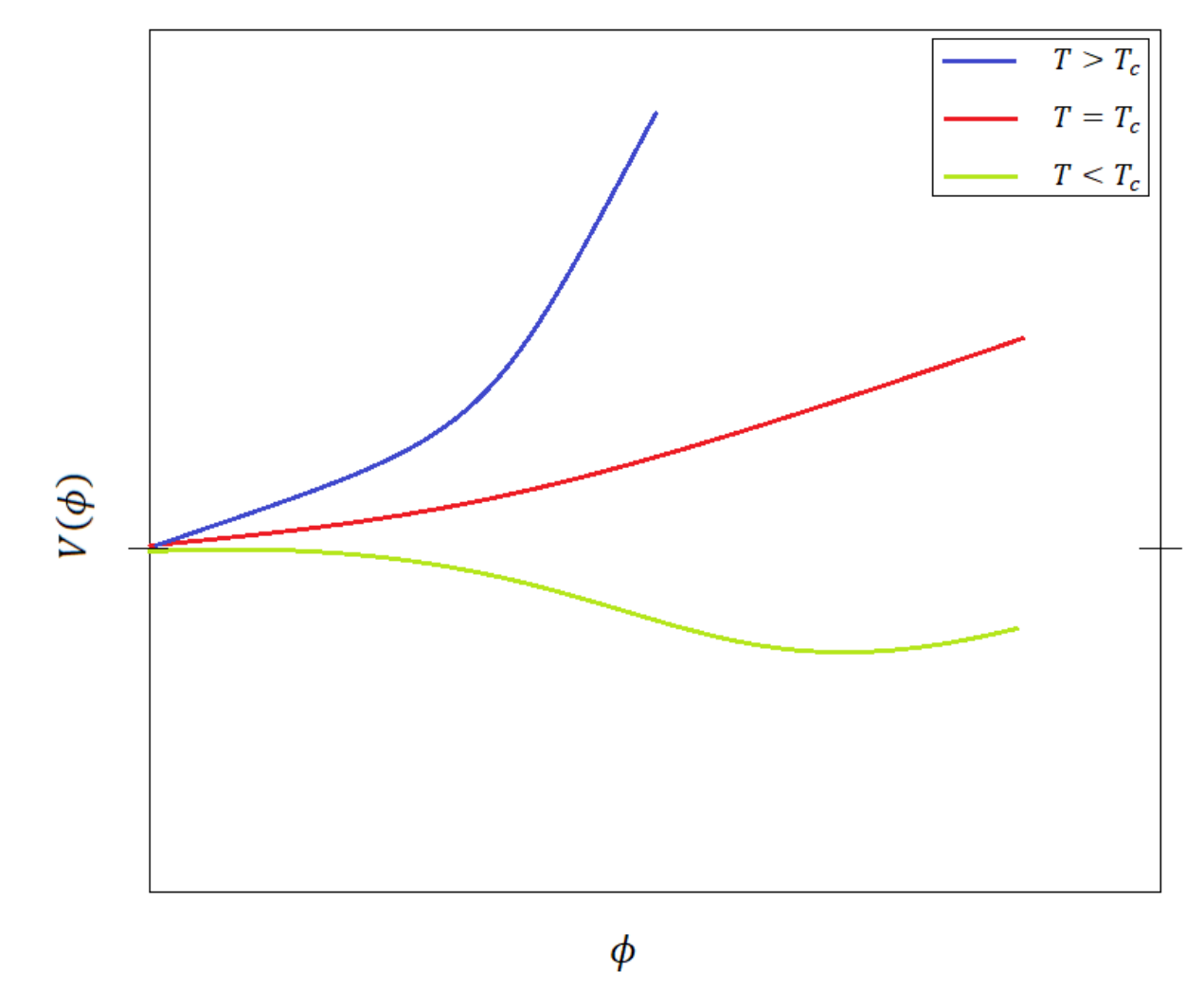}
\caption{\label{Tc variation}Scalar field vs its Potential for different values of Temperature}
\end{figure}

Following the discussion in the \ref{chap:1}, $m^2 > 0$ indicates symmetry and $m^2 < 0$ implies that the symmetry is broken. When this undergoes temperature corrections for cosmological applications, the above equation becomes:
\beq
V \left(|\phi|^2, T\right) = \left(-m^2 + \alpha T^2 \right)|\phi|^2 + \lambda_{\phi} |\phi|^4
\eeq
where, $\alpha$ is a factor containing coupling relevant constants.
It can be seen from here that the critical temperature is defined at:
\beq
T_c^2 = -m^2/\alpha
\eeq
Symmetry exists for temperatures greater $T_c$ and the symmetry is broken otherwise. Graphically, this is represented by figure \ref{Tc variation}.

Not only are the phase transitions believed to be the reason behind symmetry breaking, but it is also supposed that further appropriate phase transitions can result in the restoration of broken symmetry between the electromagnetic, strong and weak forces. This causes all the fundamental particles to be massless, facilitating the strong and weak forces analogous to the long-range electromagnetic interactions. So, it is reasonable to say that these effects can help in the understanding of the early universe, its primary properties, its evolution and other physical processes. Phase transitions could be more significant than credited for, as there is a chance that these transitions are responsible for the large-scale structure seen in the universe today. The first-order phase transition in the early universe due to bubble expansion could have triggered enormous inhomogeneities in the energy density \citep{Linde:1978px} leading to the formation of several black holes. There are more debates on the consequences of phase transitions, provided that it had occurred right after $t \approx  10^{-34}$ seconds, more commonly known as the Plank time. But this research concentrates only on the phase transitions that had possibly developed when the universe was $\sim  10^{-34}$ seconds old.

\section{Bubble growth and expansion}
First-order phase transition has numerous consequences other than the stochastic gravitational wave background. All of these phenomena are based on the theory of expanding bubbles that suggests that a first-order phase transition progresses by the nucleation, growth and coalescence of bubbles until the new phase is wholly established. The fluid, now a plasma, is treated as governed by the laws of hydrodynamics. However, the question of where these new-phase bubbles originated from in the first place is still unanswered. The closest, plausible hypothesis is the process of barrier tunnelling due to quantum fluctuations or thermodynamic instabilities. The drawback of this hypothesis is the time constraint: either of these two processes needs a substantial amount of time whose value is measured to be greater than that of the age of the universe.
 
The bubble formation is built on the rate of new phase cluster formation as a function of the quenched metastability from its state of homogeneity. This is identified as the classical nucleation theory. The key assumption of this theory is that the new phase bubbles grow owing only to the evaporation-condensation process \footnote{Referred from: \url{https://www.sciencedirect.com/science/article/pii/037596019500323U}}, a bubble of a single molecule is subjected to. The phase transition is thought to occur only if they nucleate as true-vacuum bubbles. If the bubble nucleation per unit volume is very less than the fourth power of the expansion rate, the nucleation process is rare and the universe is said to be trapped in the false vacuum phase. The equilibrium rate of nucleation of bubbles of the new phase obeys the Boltzmann distribution, according to the fluctuation theory. After a bubble nucleates, it starts expanding at the speed of light, growing in a comoving size. When it attains a size of approximately one Hubble radius, it ceases from expanding and then carries on to stretch conformally with the expansion of the universe, stabilizing the its volume fraction occupied in the universe to a constant value.
 
\begin{figure}
\centering
\includegraphics[width=15cm]{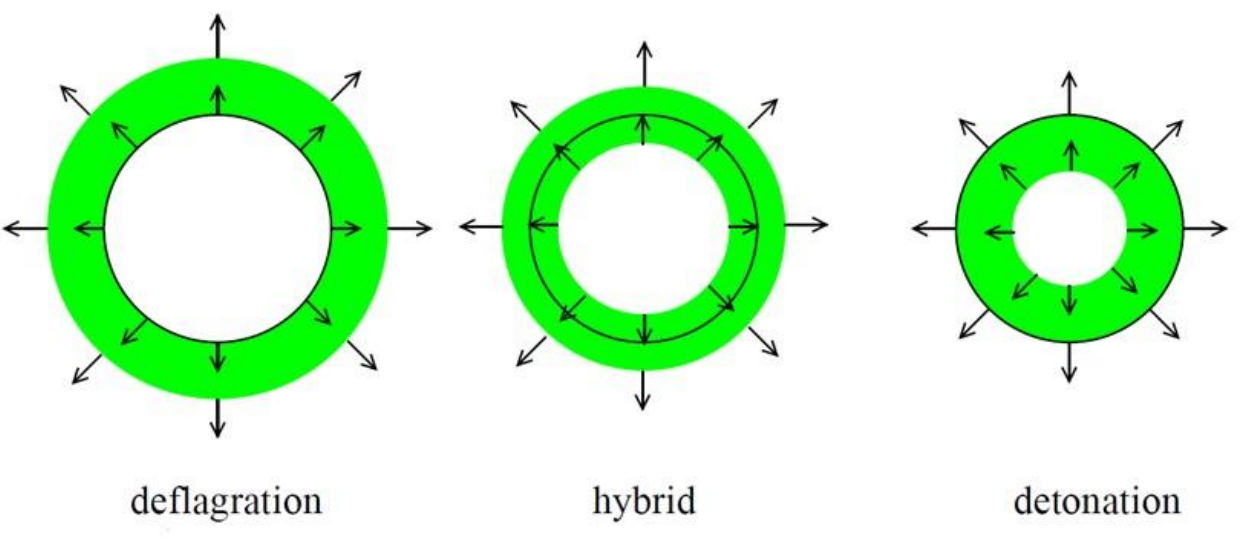}
\caption{\label{Bubble}Three cases of bubble expansion taken from \citep{Espinosa:2010hh} with the coloured area denoting the non-zero fluid velocity, hitting the bubble wall represented by the black circles}
\end{figure}

For expanding bubbles, it is necessary that the velocity profiles become zero at some point in front (upstream) or behind (downstream) the bubble wall. This is attained only through discontinuous jumps at the transition front \citep{Espinosa:2010hh}. Let $\xi_b$ be the velocity of a point in the moving fluid flow, there are three possible cases to analyze such jumps as shown in figure \ref{Bubble}.

\subsection{Detonation}
In this case, the phase transition wall has a supersonic speed, with the plasma at rest in front of the wall, and the wall hits the stationary fluid in front of the wall. In the wall-frame, symmetry exists in the fluid as it moves into the wall and is broken behind the wall. In the bubble center, or the rest-frame, the fluid behind the wall experiences a jump in its velocity to the Lorentz-transformed fluid velocity: a jump from the wall-frame to the rest-frame after the passage of the wall. This slowly dies away at some velocity less than the wall velocity, initiating a rarefaction wave in the wake of the wall, and smoothly reaches zero at the point where the fluid velocity equals $c_s$, the speed of sound.

\subsection{Deflagration}
In this case, the velocity of the fluid behind the wall is larger than that at the wall front, suggesting that the plasma rests behind the transition wall. As the transition wall moves outward, the fluid velocity in front of the wall experiences a jump to the Lorentz-transformed fluid velocity: a jump from the bubble center rest-frame to the wall-frame. Here, the velocity goes to zero due the formation of shock fronts, which has a singularity at $\xi_b = c_s$ and a strongest shock-front at the instance where the wall sweeps the plasma away along with it. This indicates that the transition is limited in a way that the bubble containing the broken phase is empty. This case of bubble expansion is common when the wall velocity is subsonic in nature.

\subsection{Hybrid}
If the wall is given to be supersonic, this case is the combination of detonations and deflagrations. It is found that the deflagrations moving at supersonic speeds are unstable due to the growth of a rarefaction wave of Jouguet type. It denotes that the wall velocity is greater than the downstream flow velocity and the upstream flow velocity is lesser than the speed of sound. Since it is a superposition of the above two cases, the downstream flow velocity is equal to the speed of sound, thus, satisfying the Chapman-Jouguet condition that a fluid lead by a shock-front reaches sonic velocity right before it ceases to be in motion. Hence, in this case, the transition wall is led by a thin shock-front and trailed by Jouguet type rarefaction wave, with the maximal flow in its front.

\section{Relativistic hydrodynamics}
As mentioned above, the bubbles are given a hydrodynamic treatment. Now is a good point to familiarize the hydrodynamic equations and conditions \citep{Rezzolla:2013dea} before proceeding further.

\subsection{Newtonian kinetic theory}
Consider a single-component system with no internal degrees of freedom, characterized by particle mass. Defining $f(t,\mathbf{\vec{x}},\mathbf{\vec{u}})$ as the distribution function for this system of indistinguishable particles at a given time $t$, such that with the probability of having a velocity $\mathbf{\vec{u}}$ at which the velocity space element $d^3 u$ exists and position $\mathbf{\vec{x}}$ at which the coordinate volume element $d^3 x$ exists, the total number of particles $N$ in the system is given by:
\beq
N = \int_{-\infty}^{\infty} d^3 x = \int_{-\infty}^{\infty} d^3 u f(t,\vec{x},\vec{u}) = \int f(t,\vec{x},\vec{u})  d^3 x  d^3 u
\eeq
Let the uniform and stationary equilibrium distribution be denoted by $f_0(\mathbf{\vec u})$. The H-theorem, which is important in the kinetic theory of fluids, puts a condition in place for the the equilibrium distribution's functional form that can be written as:
\beq
\frac{\partial{f_0}}{\partial t} = 0
\eeq
For a simple (classical) mono-atomic fluid undergoing collision, the three quantities --- particle number, total momentum, and kinetic energy --- are conserved during collision, and the equilibrium distribution is formulated as:
\beq
f_0(\mathbf{\vec u}) = C e^{-A(\mathbf{\vec u} - \mathbf{\vec {u_0}})^2}
\eeq

\subsubsection{The Maxwell-Boltzmann equilibrium distribution}
The equilibrium distribution function of a system with a single component and negligible effects from external forces is given by the Maxwell-Boltzmann distribution. Exploiting the results of H-theorem and using the following integral identities to determine the unknown coefficients:
\beq
\int_{-\infty}^{\infty} e^{-A(\vec u)^2} d^3 u = \frac{\pi^{3/2}}{A^{3/2}};
\int_{-\infty}^{\infty} \vec u e^{-A(\vec u)^2} d^3 u = 0;
\int_{-\infty}^{\infty} \vec {u^2} e^{-A(\vec u)^2} d^3 u = \frac{3}{2} \frac{\pi^{3/2}}{A^{5/2}}
\label{id2.45}
\eeq
\beq
\Rightarrow A = \frac{3}{4\epsilon};\quad C =  n \left(\frac{3}{4\pi \epsilon}\right)^{3/2}
\eeq

Since classical mono-atomic fluid obeys the laws:
\beq
\epsilon = \frac{3}{2}  \frac{k_B T}{m};\quad p = n k_B T
\eeq
where $n$ is the particle number density, $k_B$ is the Boltzmann constant, and $T$ is the temperature.
Therefore, the Maxwell-Boltzmann distribution can be written as:
\beq
f_0(t,\mathbf{\vec x},\mathbf{\vec u}) = n(t,\mathbf{\vec x}) \left(\frac{m}{2\pi k_BT(t,\mathbf{\vec x)}}\right)^{3/2} \exp \left (-\frac {m(\mathbf{\vec u}(t,\mathbf{\vec x})-\mathbf{\vec v})^2}{2k_BT(t,\mathbf{\vec x})}\right)
\eeq
This is also known as the local Maxwellian, i.e. when the thermodynamic quantities $n$, $T$ and $\vec u$ are independent functions of time and space; else, the distribution function becomes the absolute Maxwelllian, $f_0(\mathbf{\vec u})$. Computing the velocity variance by extending the average (referring to the definition of mean macroscopic velocity):
\beq
\mathbf{\vec v}_{rms}^2 := \langle \mathbf{\vec u}^2 \rangle = \frac {1}{n} \int \mathbf{\vec u}^2 f d^3 u
\eeq
also known as the root mean square velocity. Using \ref{id2.45} and $f_0$, a kinetic expression can be obtained for $T$ in terms of the system's microscopic and statistical properties as:
\beq
\langle \mathbf{\vec u}^2 \rangle = \frac{3k_BT}{m}- {\langle \mathbf{\vec u} \rangle}^2 = \frac{3k_BT}{m}- \mathbf{\vec v}^2
\eeq
\beq
\Rightarrow T=\frac{m}{3k_B}\left(\langle \mathbf{\vec u}^2\rangle- {\langle \mathbf{\vec u} \rangle}^2\right)=\frac{m}{3k_B}\left(\mathbf{\vec v}_{rms}^2- \mathbf{\vec v}^2\right)
\eeq

If the moment equations --- continuity equation, conservation of momentum equation, and conservation of internal energy equation --- have a distribution function that can be described by the Maxwell-Boltzmann distribution, then the system is said to have a zero-order approximation. Two underlying assumptions to the zero-order approximation are that the fluid under consideration is a collisional fluid (i.e. Knudsen number, ratio of mean free path of the particles to their characteristic length, $K_n << 1$) and that the system possesses local thermodynamic equilibrium (i.e. system attaining local equilibrium with approximations derived from the local Maxwellian on a time scale of the order evaluated by the ratio of mean free path length to particle velocity). So, the moment equations are:
\beq
\frac {\partial p}{\partial t} + \vec\nabla \cdot (\rho\mathbf{\vec v})=0
\eeq
\beq
\frac{\partial \mathbf{\vec v}}{\partial t} + (\mathbf{\vec v} \cdot \vec \nabla)\mathbf{\vec v} = -\frac{1}{\rho}\vec\nabla p + \frac {\mathbf{\vec F}}{m}
\eeq
\beq
\rho \left(\frac{\partial \epsilon}{\partial t} + \mathbf{\vec v} \cdot \vec \nabla \epsilon \right) + p \vec \nabla \cdot \mathbf{\vec v} = 0
\eeq
Defining $\epsilon_N$ as the total Newtonian energy density which is the quantity proportional to the contributions of internal energy and kinetic energy:
\beq
\epsilon_N := \rho \epsilon + \frac{1}{2} \rho \mathbf{\vec v}^2
\eeq
Multiplying the momentum equation by $\mathbf{\vec v}$ and adding it to the energy equation, the conservation equation for the total energy density is obtained:
\beq
\frac {\partial}{\partial t} \left(\frac{1}{2}\rho\mathbf{\vec v}^2 + \rho\epsilon \right) + \vec\nabla \cdot \left[\left(\frac{1}{2}\rho\mathbf{\vec v}^2 + \rho\epsilon + p \right) \mathbf{\vec v} \right] = \frac{\rho}{m} \mathbf{\vec F} \cdot \mathbf{\vec v}
\eeq
The term $(\frac{1}{2}\rho\mathbf{\vec v}^2 + \rho\epsilon + p) \mathbf{\vec v}$ is the Newtonian energy-flux-density vector. The fluids described by the above equations are defined as perfect fluids in the zero-order approximation given by the absolute Maxwellian equilibrium distribution. Perfect fluids are such that the viscous effects and heat fluxes are null and the pressure tensor is diagonal.

\subsubsection{Relativistic kinetic theory}
\underline{Relativistic Boltzmann equation:} The four-momentum of a particle with a rest-mass $m$ and spacetime coordinates $\vec x$ is normalized such that $p^\mu p_\mu = -m^2c^2$. In the six-dimensional phase space, the Newtonian framework describes the distribution function $f$ as:
\beq
f d^3x d^3p = f dx^1 dx^2 dx^3 dp^1 dp^2 dp^3
\eeq
Though $d^3x$ does not carry the characteristic of Lorentz invariance, the product of $d^3x d^3p$ is a Lorentz invariant. As the number of particles existing in the volume element must be concurred by any two observers, $f$ must inevitably be a Lorentz invariant as well:
\beq
f^\prime(\mathbf{x}^\prime,\mathbf{u}^\prime) d^3x^\prime d^3p^\prime = f(\mathbf{x},\mathbf{u})d^3x d^3p
\eeq
Thus, the relativistic Boltzmann equation is written as:
\beq
p^\mu \frac{\partial f}{\partial x^\mu} + m \frac {\partial (F^\mu f)}{\partial p^\mu} = \Pi(f)
\eeq
where $F^\mu$ is the four-force experienced by a particle and the scalar $\Pi$ is the generalized collision integral in the relativistic context given by:
\beq
\Pi(f) := \left(\frac {\partial f}{\partial t}\right)_{coll}
\eeq

\subsubsection{Relativistic transport fluxes}
The relativistic transport fluxes $\phi^\mu(\textbf{G})$ of a tensor $\textbf{G}$ of rank $k$ is given by:
\beq
\phi^{\mu\alpha_1...\alpha_k}(\textbf{G}) = \int G^{\alpha_1...\alpha_k}p^\mu f \frac{d^3p}{p^0}
\eeq
Setting $k=0$ and $\textbf{G} := c\mathbf{1}$ where $\mathbf{1}$ is the unit tensor, the above equation yields the first moment of $f$, a.k.a. number density current four-vector:
\beq
N^\mu := c \int p^\mu f \frac{d^3p}{p^0}
\eeq
This can be extended to define the rest-mass density current as:
\beq
J^\mu := mN^\mu = mc \int p^\mu f \frac{d^3p}{p^0}
\eeq
Here, $k$ is set to unity and $\textbf{G} := c\textbf{p}$; the relativistic transport flux equation yields the second moment of $f$, which is known as energy-momentum tensor or also called as the stress-energy tensor:
\beq
T^{\mu\nu} := c \int p^\mu p^\nu f \frac{d^3p}{p^0}
\eeq
Finally, setting $k=2$ and $\textbf{G} := c\textbf{p}\otimes\textbf{p}$, the relativistic transport flux equation yields the third moment of the distribution function:
\beq
F^{\mu\nu\sigma} := c \int p^\mu p^\nu p^\sigma f \frac{d^3p}{p^0}
\eeq
When the energy-momentum tensor is contracted with the four-velocity of the fluid, the energy density of the fluid as measured by an observer co-moving with it can be obtained from:
\beq
T_{\mu\nu}u^\mu u^\nu = \frac{1}{m^2c^2} T_{\mu\nu}p^\mu p^\nu
\eeq
Defining $\xi$ as a generic four-vector that is time-like and directed towards the future, then the following three conditions are placed:
\begin{enumerate}
\item \textit{Weak energy condition} states that the energy density in the above equation shall be positive for classical matter of any type considered reasonably:
\beq
T_{\mu\nu}\xi^\mu\xi^\nu \geq 0 \label{weakEcond}
\eeq
\item \textit{Strong energy condition} requires that the condition is stronger, though completely independent of the weak energy condition, as in:
\beq
T_{\mu\nu}\xi^\mu\xi^\nu \geq -\frac{1}{2}T^\alpha_\alpha
\eeq
\item \textit{Dominant energy condition} requires that the quantity given by $-T_{\mu\nu}\xi^\mu$ is a time-like and future-directed or a null four-vector:
\beq
-T_{\mu\nu}\xi^\mu = Au_\nu + Bk_\nu
\eeq
where $A$ and $B$ are positive numbers and $k$ is a light-like vector.
\end{enumerate}
Also, the Einstein equations, when rewritten as follows with the strong energy condition, prove the equivalence between the strong energy condition and the positivity condition for the Ricci tensor ($R_{\mu\nu}u^\mu u^\nu\geq 0$), expressed mathematically as:
\beq
\label{Eins}
R_{\mu\nu}u^\mu u^\nu = \frac{8\pi G}{c^4} \left(T_{\mu\nu} - \frac{1}{2}Tg_{\mu\nu}\right)u^\mu u^\nu = \frac{8\pi G}{c^4} \left(T_{\mu\nu} u^\mu u^\nu + \frac{1}{2}T^\alpha_\alpha\right)
\eeq

\subsubsection{General relativistic hydrodynamic equations} The rest-mass density current and the energy-momentum tensor defined above follows from the the relativistic H-theorem and can be expressed as the continuity equation and the energy-momentum conservation equation respectively, as in Newtonian kinetic theory:
\beq
\frac{\partial J^\mu}{\partial x^\mu} = mc \frac{\partial}{\partial x^\mu} \int p^\mu f \frac{d^3p}{p^0} = 0
\eeq
\beq
\frac{\partial T^{\mu\nu}}{\partial x^\mu} = c \frac{\partial}{\partial x^\mu} \int p^\mu p^\nu f \frac{d^3p}{p^0} = c \int m F^\nu f \frac{d^3p}{p^0}
\eeq
These equations are useful only if the distribution function $f$ is known. But a fully general-relativistic expression for the hydrodynamic equations can be obtained by exploiting the tensor formulation. For a given ($\Sigma$) which is defined as a closed three-surface, the conservations of rest-mass and energy-momentum imply the vanishing of the total net fluxes across it; and since $J^\mu$ and $T^{\mu\nu}$ represent the fluxes of rest-mass and four-momentum respectively, the condition for a flat spacetime with a unit normal $l_\mu$ to the three-dimensional surface $\Sigma$ becomes:
\beq
\int_\Sigma J^\mu l_\mu d^3x = 0
\eeq
\beq
\int_\Sigma T^{\mu\nu} l_\mu d^3x = 0
\eeq
Transforming these equations to volume integrals using Gauss divergence theorem and noting that these conservation laws hold only if:
\beq
\nabla_\mu J^\mu = 0
\eeq
\beq
\nabla_\nu T^{\mu\nu} = 0
\eeq
are satisfied, the general-relativistic hydrodynamic equations for a generic fluid can be derived. The above differential conservation laws denote the conservation of the physical quantities rest-mass, energy and linear momentum. The co-variance of the tensor formulation of these equations imply that they take the same form in a general curved spacetime.

\subsubsection{Relativistic equilibrium distributions}
When the quantum effects are neglected, the thermal effects in the relativistic fluid are important, i.e. pressure does not depend on temperature, and the fluid is referred to as non-degenerate fluid --- in this case of fluids, the equilibrium distribution when the external forces are absent is:
\beq
f_0(\textbf{p}) = \left(\frac {g_{\mathrm{\bar{s}}}}{h^3_P}\right) \frac{1}{\exp (-\alpha_f - c p^\mu u_\mu / k_B T)}
\eeq
where $p^\mu = mcU^\mu$ is the four-momentum of the particles, $u^\mu$ is the four-velocity of the fluid such that $p^\mu u_\mu \neq -mc$, and $g_{\mathrm{\bar{s}}}$ is the degeneracy factor that accounts for its internal degrees of freedom that is defined by its rest-mass $m$ and particle spin $\mathrm{\bar{s}}$ as:
\beq
g_{\mathrm{\bar{s}}} = 
\begin{cases} 
    2\mathrm{\bar{s}}+1, & m \neq 0 \\
    2\mathrm{\bar{s}}, & m=0 
\end{cases}
\eeq
For a rest-mass density $\rho := nm$, total energy density $e:=\rho \left(c^2 + \epsilon \right)$ (hence, specific energy $e/\rho = c^2 + \epsilon$) and specific entropy $s:= S/Nm$, the quantity $\alpha_f$ is known as fugacity in thermodynamics (which is a measure of the degree of the fluid's degeneracy) is given by:
\beq
\alpha_f \equiv \frac{m}{k_B T} \left( \frac{e+p}{\rho}-Ts \right)
\eeq
This is called Maxwell-J$\mathrm{\ddot u}$ttner distribution function that generalizes the Maxwell-Boltzmann distribution in the relativistic framework.

In case of non-degenerate relativistic fluid with a co-moving (with the fluid) reference frame $u_\mu = (-1, 0, 0, 0)$ and Lorentz factor with respect to the fluid frame $w = \left(1-V^i V_i/c^2 \right)^{-1/2}$ so that $-c p^\mu u_\mu = wmc^2$ and $p^2 = m^2 c^2 (w^2 -1)$, and depending on the Lorentz factor, $w$ rather than $p$ giving a converted distribution $f_0(w) dw \equiv f_0(p) dp \times 4\pi p^2 dp$, the special case expression reduces to:
\beq
f_0(w) = \frac {\zeta_c n}{K_2(\zeta_c)} w(w^2-1)^{1/2} \exp{(-w\zeta_c)}
\eeq
Relativistic coldness $\zeta_c \equiv {mc^2}/{k_B T}$ measures the fluid's thermal energy and $K_2(\zeta_c)$ is a modified Bessel function of second kind.
When quantum effects are considered, the volume element becomes $d^3x d^3p/h_P^3$, where $h_P$ is the Planck constant. For a given occupation number and spin, the particle can either be a boson or a fermion. For $-c u_\mu p^\mu = E = wmc^2$, the degenerate fluid can be expressed for both distinct equilibrium distributions as:
\beq
f_0(\mathbf{p}) = \left(\frac{g_{\mathrm{\bar{s}}}}{h_P^3}\right) \frac{1}{\exp[(E-\mu)/k_B T] \pm 1}
\eeq
Fermions take the positive sign and the bosons take the negative sign.

Thus, relativistic perfect fluids are the fluids which --- at the zero-order approximation of the equilibrium distributions --- can be expressed by the moment equations, having zero viscous effects, no heat fluxes, and a diagonal pressure tensor.

\subsubsection{Some thermodynamics}
The first law of thermodynamics for a fluid at an absolute temperature $T$ with entropy $S$, pressure $p$, occupying a volume $V$, and a single type of particle with its number conserved:
\beq
dU = T dS - p dV
\eeq
where, the internal energy $U$ defined (for density $\rho$ and specific energy $\epsilon$) as $U = \rho \epsilon V$ is exact i.e. path-independent inter-state energy changes.

The second law of thermodynamics for an arbitrary space-like surface with a total entropy $S$ is equivalent to the condition of non-decreasing entropy flux, defined as the the rate of change of entropy within the system due to entropy transfer, given by:
\beq
\nabla_\mu S^\mu \geq 0
\eeq
This is known as the maximum entropy principle.  It implies the relation between the microscopic nature of entropy and the properties of collision integral. From this law, it can be shown that along fluid lines, the specific entropy obeys conservation --- hence, a perfect fluid would be an adiabatic fluid.

\subsection{Equations of state}
The equilibrium function and moment equations are sufficient to calculate all the thermodynamic variables of a fluid. But the distribution function is highly complicated to be determined using the (collisional) Boltzmann equation. It is convenient to establish a relation between the quantities such as the energy-momentum tensor to other hydrodynamic variables, called as equations of state. Defining dimensionless momentum as:
\beq
\cosh x \equiv \frac{p^0}{mc}\quad
\eeq
Normalized as:
\beq
p^\mu p_mu = -mc^2 = (p^0)^2 - |\mathbf{\vec{p}}|^2
\eeq
\beq
\Rightarrow |\mathbf{\vec{p}}|^2 = m^2 c^2 (\cosh^2 x - 1) = m^2 c^2 \sinh^2 \quad \mathrm{Or:} \quad \frac{d |\mathbf{\vec{p}}|}{p^0} = dx
\eeq
Let $\chi$ represent a particular combination of momentum. Distribution function is now:
\beq
\int \chi f \frac{d^3p}{p^0} = 4\pi \frac{g_{\mathrm{\bar{s}}}}{h_P^3} m^2 c^2 \int_0^\infty dx \frac{\chi \sinh^2 x}{\exp (-\alpha_f + \zeta_c \cosh x) \pm 1}
\eeq
Irrespective of the complexity of the integrals, the fugacity (thermodynamic property which accounts for the pressure in real gas from ideal gas relations) and relativistic coldness (a measure of fluid thermal energy proportional to the reciprocal of temperature) factors highlight the behaviour of the fluid. If $\alpha_f \ll 1$, fluid is non-degenerate due to unimportant quantum effects, whereas, if $\alpha_f \gg 1$, fluid is degenerate due to the quantum effects. Similarly, $\zeta_c \ll 1$ implies that the thermal energy is very high compared to the rest-mass energy and the fluid is relativistic, while, $\zeta_c \gg 1$ implies that the fluid is non-relativistic. When $p$ depends on the total energy density, $e$, and specific entropy, $s$, then a thermodynamic identity for the speed of sound exists as:
\beq
c_s^2 = \left(\frac{\partial p}{\partial e}\right)_s
\eeq

\subsubsection{Ultra-relativistic fluid}
When the physical conditions of a fluid are limited to the vanishing of the coldness $\zeta_c \ll 1$, the equations of state for the ultra-relativistic (perfect) fluids pop out. Defining an integral variable change such that $z \equiv \zeta_c \cosh x$ and imposing the condition of internal energy being larger than the rest-mass energy (e.g., neutrinos), it can be assumed that $(z/\zeta_c)^2 \approx \cosh^2x - 1$, the equation for $p$ for a degenerate ultra-relativistic fluid becomes:
\beq
p = \frac{1}{3} 4\pi \frac{g_{\mathrm{\bar{s}}}}{h_P^3} c \left(\frac{k_B T}{c}\right)^4 \int_0^\infty dz \frac{z^3}{\exp (-\alpha_f + z) \pm 1}
\eeq
This gives the following pair of important properties pertaining to ultra-relativistic fluids:
\beq
p = \frac{1}{3}e \quad\quad;\quad\quad c_s^2 = \frac{1}{3}
\eeq

\subsection{Kinematic properties of fluid}
In a co-ordinate system $x^\mu = (t, x^i)$, the contravariant components of the four-velocity, $u$, for a proper time $\tau$ when the observer is comoving with the fluid, is expressed as:
\beq
u^\mu := \frac{d x^\mu}{d\tau}
\eeq
This brings about the kinematic fluid four-acceleration:
\beq
a^\mu := u^\nu \nabla_\nu u^\mu
\eeq
Recalling the normalization condition:
\beq
\mathbf{u \cdot u} = u^\mu u_\mu = -1 
\eeq
and the condition of orthogonality:
\beq
\mathbf{a \cdot u} = a^\mu u_\mu = 0
\eeq
the following identity is derived:
\beq
u^\mu \nabla_\nu u_\mu = 0
\eeq
where, the components of four-velocity for a co-ordinate time $t$ is written as $u^\mu = u^0 (1, v^i)$.

\subsection{Energy-momentum of perfect fluids}
Newtonian kinetic theory summarized that the equilibrium distribution function characterizes an energy flux that vanishes and an isotropic pressure tensor (perfect fluid properties that should be preserved in a relativistic framework) in terms of the locally averaged flow velocity of the fluid element. If a comoving observer in the local rest frame $\check{\bullet}$ measures a zero matter flux in any direction, the components of the rest-mass density current are:
\beq
J^{\check{\mu}} = (\rho,0,0,0)
\eeq
in the $\check{\mu}$ direction. Note that the energy-momentum tensor $T^{\check{\mu} \check{\nu}}$ denotes the $\check{\mu}$-momentum flux in the $\check{\nu}$ direction. $T^{\check{\mu} \check{\nu}}$ is required to be a symmetric tensor, for the average relativistic fluid particle energy $E = \langle \check{p}^0 \rangle$; its components can be determined from these conditions:

\noindent Total energy density is equal to internal energy plus rest-mass energy:
\beq
T^{\check{0}\check{0}} = e = n \langle \check{p}^0 \rangle
\eeq
Energy density flux in the $\check{i}^{\mathrm{th}}$ direction:
\beq
T^{\check{0}\check{i}} = 0
\eeq
$\check{i}$-momentum density flux in the $\check{0}^{\mathrm{th}}$ direction:
\beq
T^{\check{i}\check{0}} = 0
\eeq
$\check{i}$-momentum density flux (for $i \neq j$) in the $\check{j}^{\mathrm{th}}$ direction:
\beq
T^{\check{i}\check{j}}=T^{\check{j}\check{i}} = 0 
\eeq
$\check{j}^{\mathrm{th}}$ component (for $i = j$) of momentum density flux in the $\check{i}^\mathrm{th}$ direction:
\beq
T^{\check{j}\check{i}} = p
\eeq
If $u^{\check{\mu}} := (1, 0, 0, 0)$ is the fluid four-velocity, the rest-mass density current is rewritten as:
\beq
J^{{\mu}} = \rho u^{{\mu}}
\eeq
in any other frame. Similarly, the energy-momentum tensor can be generalized as:
\beq
T^{\mu\nu} = (e+p) u^\mu u^\nu + p g^{\mu\nu} = \begin{pmatrix}
e & 0 & 0 & 0\\
0 & p & 0 & 0\\
0 & 0 & p & 0\\
0 & 0 & 0 & p
\end{pmatrix}
\eeq
where $g^{\mu\nu}$ is the metric tensor.
A fluid behaves like dust when it is pressure-less --- the enthalpy density defined as $\mathrm{w} := e+p$ for this case ($p=0$) reduces the above equation to:
\beq
T^{\mu\nu} = \mathrm{w} u^\mu u^\nu 
\eeq

\section{Kolmogorov turbulence}
\label{KT}
The hydrodynamic interactions of the bubbles become more prominent after symmetry breaking due to their expansion and collisions, creating turbulence. This turbulence is a potential source of the stochastic gravitational wave background. Since the influence of turbulence is becomes crucial after the phase transition commences, it is necessary to adopt a model which can account for some of its significant properties. Identifying the Kolmogorov theory here explains a few parameters. This is an asymptotic theory that has been proven to work perfectly at high Reynolds number. Reynolds number is a dimensionless quantity that predicts if the fluid flow is steady or turbulent.

\subsection{Eddies}
When a fluid undergoing a turbulent flow encounters a solid obstacle, it has a tendency to surge past the object in such a way that there is a void created at the solid-fluid interface. This void gives rise to circular, reverse current called eddies that acts as a small, swirling perturbation that continues to grow through the turbulent fluid over time.
 
Kolmogorov theory helps in understanding the energy mechanisms of the eddies like the transfer of energy from large to small eddies, dissipation according to their size, etc. Any form of perturbation, such as an obstacle, an external force or an injected kinetic energy can give rise to eddies.
 
Initially, large eddies form in the turbulent fluid, gathering energy from the first encounter of the disturbance. These large eddies are highly unstable and end up breaking and transferring their energy into smaller eddies. The process of energy cascade is observed while the smaller eddies collapse into even smaller eddies and transfer its energy, which further breaks down and transfers energy into yet smaller eddies.
 
This process of collapsing of eddies and transferring of energy in a successive manner continues until eddy motion becomes stable due to the sufficient reduction in Reynolds number and the flow viscosity becomes effective to dissipate the acquired energy in the system. According to Kolmogorov, this was a chaotic scale reduction process that have no directional bias, suggesting that the energy cascade is one-way: from large to small eddies. He formulated his first hypothesis from here, known as Kolmogorov's hypothesis of local isotropy, which states that:
 \begin{quote}
    \centering
    \textit{``At sufficiently high Reynolds numbers, the small-scale turbulent motions are statistically isotropic.''}
\end{quote}

\noindent He further argued that the eddy geometry is also lost during the energy cascade. This gave rise to the Kolmogorov's first similarity hypothesis:

\begin{quote}
    \centering
    \textit{``In every turbulent flow at sufficiently high Reynolds number, the statistics of the small scale motions have a universal form that is uniquely determined by rate of energy dissipation and kinematic viscosity.''}
\end{quote}

The size range specified by the small scale motions of the eddies is known as the universal equilibrium range, implying that the small eddies quickly adapt to changing energy conditions and maintain a dynamic equilibrium. The scales at which the energy is dissipated indicates the smallest eddies in the flow. Given the rate of dissipation ($\epsilon$) and kinematic viscosity ($\nu$), this scale yields the following parameters. The length scale that corresponds to the smallest eddies in the flow is called the Kolmogorov micro-scale.
\newline
Length scale:
\beq
\lambda = \left(\frac{\nu^3}{\epsilon}\right)^{1/4}
\eeq
Velocity scale:
\beq
u_{\lambda} = (\epsilon \nu)^{1/4}
\eeq
Time scale:
\beq
\tau_{\lambda} = \left(\frac{\nu}{\epsilon} \right)^{1/2}
\eeq
This theory is limited due to its assumption that the energy cascade is uni-directional, since it has been experimentally proved that the process of backscatter, the reverse energy transfer (from smaller to larger eddies) is also possible, however weak. But since the early universe had a huge value of Reynolds number, this limitation is ignored and employed in this study.

\chapter{Fluid interactions}
Bubbles start nucleating during the phase transition from a stable phase to the quenched metastable phase. The large vacuum expectation value gained by the scalar field breaks the symmetry spontaneously. Whilst the bubbles collide and enter into the broken phase, depending on the type of expansion (detonation, deflagration or hybrid), shock waves and/or rarefaction waves are produced. This might have resulted in the production and propagation of sound waves which would exist even after the extinction of their stable phase boundaries.
 
In this outlook, given an infinitesimal displacement vector $dx$, its infinitesimal line element is written as:
\beq
ds^2 = -dt^2 + (\delta_{ij} + h_{ij}) dx^i dx^j
\eeq
where, $h_{ij}$ is the metric perturbation The transverse traceless part of the energy-momentum tensor acts as the source to the metric tensor and can be related as:
\beq
\label{3.2}
\ddot{h}_{ij} - \nabla^2 h_{ij} = 16\pi G.\Pi_{ij}
\eeq
The quantity $\Pi_{ij}$ is contributed by both the fluid and the scalar field as:
\beq
\label{Pif}
\Pi_{ij}^f = [(e+p) \gm^2 v_i v_j + p \dl_{ij}]^{TT}
\eeq
\beq
\label{Pis}
\Pi_{ij}^\phi = [\dau_i \phi \dau_j \phi - \frac{1}{2} (\dau\phi)^2\dl_{ij}]^{TT}
\eeq
The general solution for the metric tensor and, thus, the radiation equation is given by:
\beq
h_{ij} (\textbf{x},t) = 4G \int d^3x^\prime \frac{\Pi_{ij} (\textbf{x}^\prime, t - |\textbf{x} - \textbf{x}^\prime|)}{|\textbf{x} - \textbf{x}^\prime|}
\eeq
Section \ref{gov eq} elaborates the above equation further along with the finding of its particular solution and the expected gravitational wave spectra. The determination of the source of metric perturbations $\Pi_{ij}$, after a little more insight and definition of some quantities in following chapters, is done in detail in \ref{apxa}.

\section{Stationary turbulence model}
\label{GKK}
According to the Kolmogorov turbulence discussed in Chapter \ref{chap:lits}, the turbulent kinetic energy generated over a length scale cascades from larger to smaller scales. This is called forward cascade. To model for cosmological turbulence, the energy density of the vacuum is assumed to be converted into the kinetic energy of the plasma over a characteristic length scale in the early universe. Since only forward cascade is taken into account, the turbulent energy thermalizes and the cascade damps to dissipation.

\subsection{Formalism}
\label{gov eq}
The radiation equation for the phase transition induced turbulence follows from equation \ref{Eins} for an adiabatic environment as:
\beq
\label{eq. 4.1}
\frac {\partial^2}{\partial t^2} h_{ij}(\textbf{x},t) - \nabla^2 h_{ij}(\textbf{x},t) = 16 \pi G S_{ij} (\textbf{x},t)
\eeq
where, $h_{ij} (\textbf{x},t)$ is the metric tensor, $S_{ij} (\textbf{x},t)$ is the transverse traceless part (refer equations \ref{Pif} and \ref{Pis}) of the energy-momentum tensor $T_{ij}(\textbf{x},t)$ and $t$ is the conformal time of the physical space.
 
After Fourier transforming the equation \ref{eq. 4.1}, the particular solution of the metric perturbation following from the equation \ref{3.2} is obtained as:
\beq
h_{ij}(\textbf{k},t) = (16\pi G) \Lambda_{ij,kl}((\textbf{k}) \int_0^t dt\p \frac{\sin[k(t-t\p)]}{k}T_{kl}(\textbf{k},t\p)
\eeq
The energy-momentum is assumed to have originated from a random, isotropic process:
\beq
\Lambda_{ij,kl} ((\textbf{k}) \equiv P_{ik}((\textbf{k})P_{jl}((\textbf{k}) - \frac{1}{2} P_{ij}(\textbf{k})P_{kl}(\textbf{k})
\eeq
is the fluid shear stress and the projector is defined as:
\beq
P_{ij}(\textbf{k}) = \dl_{ij} - \hat k_i \hat k_j
\eeq
This facilitates in the computation of the necessary gravitational wave parameters. Starting off with the energy density of the gravitational waves averaged over a long time and many wavelengths:
\beq
\rho_{GW} = \frac{1}{32\pi G} \overline{\dot{h}_{ij}(x) \dot{h}_{ij}(x)}
\eeq
where the time derivative of metric perturbations is given by:
\beq
\overline{\dot{h}_{ij}(x) \dot{h}_{ij}(x) } = P_{\dot{h}}(\mathbf{k}, t)(2\pi)^3 \delta(\mathbf{k}+\mathbf{k\p})
\eeq
Here, $P_{\dot{h}}(\mathbf{k}, t)$ is its spectral density. When the turbulence is localized and bounded in a region centered around $\textbf{x}=0$, the term $|\textbf{x} - \textbf{x}^\prime|$ reduces to $|\textbf{x}|$. Now, the energy density is seen to be associated with the spectral density as:
\beq
\rho_{GW} = \frac{1}{32\pi G} \frac{1}{2\pi^2} \int dk. k^2 P_{\dot{h}}(k)
\eeq
In terms of $\mathcal{P}_{\dot{h}(k)} := (k^3/2\pi^2) P_{\dot{h}}(k)$, which is the $\dot{h}$ power spectrum:
\beq
\mathcal{P}_{GW}(k) \equiv \frac{d\Omega_{GW}}{d\ln k} = \frac{1}{12H^2} \mathcal{P}_{\dot{h}}(k)
\eeq
For unequal times, the necessary corrections to the shear stress corresponds to a spectral density, averaged over a number of oscillations with a wavenumber $k$, as:
\beq
P_{\dot{h}}(k, t) = \frac{(16\pi G)^2}{2} \int_0^t dt_1 \int_0^t dt_2 \cos(kt_1 - kt_2) \Pi^2(k, t_1, t_2)
\eeq
Since the source is homogeneous, the averaged energy-momentum correlators do not depend on space or direction. This implies isotropy, giving the following expression for the anistropic stress with a velocity field $\bar{v}_{\mathbf{q}}(t)$:
\beq
\label{Pisq}
\Pi^2 (k, t_1, t_2) = \bar{w}^2 \int \frac{d^3 q}{(2\pi)^3} \frac{q^2}{\tilde{q}^2} \left(1-\mu^2 \right)^2 G(q, t_1, t_2) G(\tilde{q}, t_1, t_2)
\eeq
where:
\beq
\mathbf{\tilde{q}} = \mathbf{q} - \mathbf{k}
\eeq
\beq
\mu = \hat{\mathbf{q}} \cdot \hat{\mathbf{k}}
\eeq
and $G(q, t_1, t_2)$ is the velocity power spectrum examined in detail in the following sections.

\section{Features of the turbulence model}
As mentioned earlier, the equation \ref{Pisq} was for the sound shell model. From this point, $\Pi$ and $\Pi^2$ are treated as interchangeable for simplicity.

\subsection{Kolmogorov hypothesis}
Kolmogorov theory examined in section \ref{KT} describes how the energy transfer from, to and between eddies of various sizes. The energy spectrum can be expressed as:
\beq
E(k)=v^2 \lambda f(k\lambda) = \nu^{5/4} \epsilon^{1/4}f(k\lambda)
\eeq
This is the mathematical form of the first similarity principle. Here, $k$ is the wavenumber, $\lambda$ is the Kolmogorov dissipation length scale, $\nu$ is the viscosity and $\epsilon$ is the dissipation rate. As Reynolds number tends to infinity, $E(k)$ becomes independent of viscosity.
 
Defining a constant $\alpha$, the second similarity principle is given by:
\beq
f(k\lambda)=\alpha(k\lambda)^{-5/3} = \alpha \nu^{-5/4}\epsilon^{5/12}k^{-5/3}
\eeq
And the energy spectrum becomes:
\beq
E(k)|_{Re\rightarrow \infty} = \alpha \epsilon^{2/3}k^{-5/3}
\eeq
Velocity fluctuations lose the directional bias as the cascade proceeds to the smallest eddies establishing local isotropy for very high Reynolds number.

\subsubsection{Stationary Kolmogorov turbulence}
The two-point velocity correlation function defined as:
\beq
R_{ij} (\mathbf{r}, \tau) \equiv \langle v_i(\textbf{x}, t) v_j(\textbf{x}+\textbf{r}, t+\tau) \rangle
\eeq
has a Fourier transform given by the spectral function $M_{ij}(\textbf{k},\tau)$ for an isotropic and homogeneous stationary turbulence and is expressed as:
\beq
M_{ij}(\textbf{k},\tau) = \frac {E_k}{4\pi k^2} \left(\delta_{ij} - \frac{k_ik_j}{k^2} \right) f(\eta_k,\tau)
\eeq
where, $\eta_k$ is the auto-correlation function and $f(\eta_k,\tau)$ is the time de-correlation function, which is negligible as $\tau \gg 1/\eta_k$.
Thus, the energy density spectrum for stationary turbulence has a power law:
\beq
E_k = C_k \epsilon^{2/3}k^{-5/3}
\eeq
The value of $k$ ranges between the wavenumbers determined by the characteristic length scale and the dissipation length scale, $C_k$ is the constant and is taken as $1$ for simplicity and $\epsilon$ is the energy dissipation rate per unit enthalpy. The auto-correlation function corresponds as:
\beq
\eta_k = \frac{1}{\sqrt{2\pi}} \epsilon^{1/3}k^{2/3}
\eeq
Adopting Kraichnan's exponential function to model for the temporal decorrelation:
\beq
\label{gkkexp}
f(\eta_k,\tau) = \exp \left(-\frac{\pi}{4}\eta_k^2 \tau^2 \right)
\eeq

\subsection{Corrections for obtaining turbulence model}
Governing equations can be listed as follows from the above discussion:
\newline
$$\left\langle v_{\tb{q}_1}^i v_{\tb{q}_2}^j \right\rangle = M^{ij}(q_1) (2\pi)^3 \delta(\tb{q}_1 + \tb{q}_2)  $$ 
$$P_\pp^{ij}(\tb{q}) = \delta^{ij} - \hat{q}^i \hat{q}^j$$
$$P_\pl^{ij}(\tb{q}) = \hat{q}^i \hat{q}^j $$
$$\Lambda^{ijkl}(\tb{k}) = P_\pp^{ik}(\tb{k}) P_\pp^{jl}(\tb{k}) - \frac{1}{2} P_\pp^{ij}(\tb{k}) P_\pp^{kl}(\tb{k})$$
$$ \Pi(k) = \bar{w}^2 \int \frac{d^3 q}{(2\pi)^3} \left [M^{ik}(q) M^{jl}(\tl{q}) + M^{il}(q) M^{jk}(\tl{q}) \right] \Lambda^{ijkl}(\tb{k})$$
$$ M^{ij}(q) = P_\pp^{ij}(\tb{q})G_\pp(q) + P_\pl^{ij}(\tb{q})G_\pl(q)$$
$$\tl{\mathbf{q}} = \tb{q} - \tb{k} $$
 $$  \mu = \hat{\mathbf{q}} \cdot \hat{\mathbf{k}} $$
 $$   \tl{\mu} = \hat{\tl{\mathbf{q}}} \cdot \hat{\mathbf{k}} $$
  $$  \tl{q}^2(1-\tl{\mu}^2) = q^2 (1-\mu^2) $$
Splitting the integrand $\Pi(k)$ as $\Pi_{[1]}$ and $\Pi_{[2]}$, and leaving out the constants:
\newline
\begin{align*}
\Pi_{[1]} &= M^{ik}(q) M^{jl}(\tl{q}) \Lambda^{ijkl}(\tb{k})\\
 \Pi_{[2]} &=M^{il}(q) M^{jk}(\tl{q}) \Lambda^{ijkl}(\tb{k})\\
\end{align*}
Evaluating:
 
\begin{align*}
\Pi_{[1]} = & \left[M^{ik}(q) M^{jl}(\tl{q})\right] \Lambda^{ijkl}(\tb{k})\\
= & \left[P_\pp^{ik}(\tb{q}) G_\pp(q) +  P_\pl^{ik}(\tb{q}) G_\pl(q)\right] \left[P_\pp^{jl}(\tl{\tb{q}}) G_\pp(\tl{q}) +  P_\pl^{jl}(\tl{\tb{q}}) G_\pl(\tl{q})\right] \Lambda^{ijkl}(\tb{k}) \\
= &\left\{ P_\pp^{ik}(\tb{q}) P_\pp^{jl}(\tl{\tb{q}}) G_\pp(q)G_\pp(\tl{q}) + P_\pp^{ik}(\tb{q}) P_\pl^{jl}(\tl{\tb{q}}) G_\pp(q) G_\pl(\tl{q})\right\} \Lambda^{ijkl}(\tb{k})\\
&+ \left\{ P_\pl^{ik}(\tb{q}) P_\pp^{jl}(\tl{\tb{q}}) G_\pl(q) G_\pp(\tl{q})+ P_\pl^{ik}(\tb{q}) P_\pl^{jl}(\tl{\tb{q}}) G_\pl(q) G_\pl(\tl{q})\right\}  \Lambda^{ijkl}(\tb{k})\\
= & \left\{\left[\delta^{ik} - \hat{q}^i \hat{q}^k \right] \left[\delta^{jl} - \hat{\tl{q}}^j \hat{\tl{q}}^l \right] G_\pp(q)G_\pp(\tl{q}) + \left[\delta^{ik} - \hat{q}^i \hat{q}^k \right]\left[\hat{\tl{q}}^j \hat{\tl{q}}^l\right] G_\pp(q) G_\pl(\tl{q})\right\} \Lambda^{ijkl}(\tb{k})\\
&+ \left\{ \left[\hat{q}^i \hat{q}^k \right] \left[\delta^{jl} - \hat{\tl{q}}^j \hat{\tl{q}}^l \right] G_\pl(q) G_\pp(\tl{q})+\left[\hat{q}^i \hat{q}^k\right] \left[\hat{\tl{q}}^j \hat{\tl{q}}^l\right] G_\pl(q) G_\pl(\tl{q})\right\} \Lambda^{ijkl}(\tb{k})\\
= & \left\{\left[\delta^{ik} \delta^{jl} - \delta^{ik} \hat{\tl{q}}^j \hat{\tl{q}}^l - \delta^{jl} \hat{q}^i \hat{q}^k +  \hat{q}^i \hat{q}^k \hat{\tl{q}}^j \hat{\tl{q}}^l \right] G_\pp(q)G_\pp(\tl{q})+\left[\hat{q}^i \hat{q}^k \hat{\tl{q}}^j \hat{\tl{q}}^l\right] G_\pl(q) G_\pl(\tl{q})\right\} \Lambda^{ijkl}(\tb{k})\\
& + \left\{\left[\delta^{ik}\hat{\tl{q}}^j \hat{\tl{q}}^l - \hat{q}^i \hat{q}^k \hat{\tl{q}}^j \hat{\tl{q}}^l\right] G_\pp(q) G_\pl(\tl{q})+ \left[\delta^{jl} \hat{q}^i \hat{q}^k - \hat{q}^i \hat{q}^k \hat{\tl{q}}^j \hat{\tl{q}}^l \right] G_\pl(q) G_\pp(\tl{q})\right\} \Lambda^{ijkl}(\tb{k})
\end{align*}
Substituting for $\Lambda^{ijkl}(\tb{k})$ and simplifying:
 
\begin{align*}
\Pi_{[1]} = & \,\, G_\pp(q)G_\pp(\tl{q})  \left[\delta^{ik} \delta^{jl} - \delta^{ik} \hat{\tl{q}}^j \hat{\tl{q}}^l - \delta^{jl} \hat{q}^i \hat{q}^k +  \hat{q}^i \hat{q}^k \hat{\tl{q}}^j \hat{\tl{q}}^l \right] \Lambda^{ijkl}(\tb{k})\\
& + G_\pp(q) G_\pl(\tl{q}) \left[\delta^{ik}\hat{\tl{q}}^j \hat{\tl{q}}^l - \hat{q}^i \hat{q}^k \hat{\tl{q}}^j \hat{\tl{q}}^l\right] \Lambda^{ijkl}(\tb{k}) \\
& + G_\pl(q) G_\pp(\tl{q}) \left[\delta^{jl} \hat{q}^i \hat{q}^k - \hat{q}^i \hat{q}^k \hat{\tl{q}}^j \hat{\tl{q}}^l \right]\Lambda^{ijkl}(\tb{k}) \\
& +  G_\pl(q) G_\pl(\tl{q}) \left[\hat{q}^i \hat{q}^k \hat{\tl{q}}^j \hat{\tl{q}}^l\right] \Lambda^{ijkl}(\tb{k}) \\
= & \,\, G_\pp(q)G_\pp(\tl{q})  \left[\delta^{ik} \delta^{jl} - \delta^{ik} \hat{\tl{q}}^j \hat{\tl{q}}^l - \delta^{jl} \hat{q}^i \hat{q}^k +  \hat{q}^i \hat{q}^k \hat{\tl{q}}^j \hat{\tl{q}}^l \right] \left[ P_\pp^{ik}(\tb{k}) P_\pp^{jl}(\tb{k}) - \frac{1}{2} P_\pp^{ij}(\tb{k}) P_\pp^{kl}(\tb{k})\right]\\
& + G_\pp(q) G_\pl(\tl{q}) \left[\delta^{ik}\hat{\tl{q}}^j \hat{\tl{q}}^l - \hat{q}^i \hat{q}^k \hat{\tl{q}}^j \hat{\tl{q}}^l\right] \left[ P_\pp^{ik}(\tb{k}) P_\pp^{jl}(\tb{k}) - \frac{1}{2} P_\pp^{ij}(\tb{k}) P_\pp^{kl}(\tb{k})\right] \\
& + G_\pl(q) G_\pp(\tl{q}) \left[\delta^{jl} \hat{q}^i \hat{q}^k - \hat{q}^i \hat{q}^k \hat{\tl{q}}^j \hat{\tl{q}}^l \right] \left[ P_\pp^{ik}(\tb{k}) P_\pp^{jl}(\tb{k}) - \frac{1}{2} P_\pp^{ij}(\tb{k}) P_\pp^{kl}(\tb{k})\right] \\
& +  G_\pl(q) G_\pl(\tl{q}) \left[\hat{q}^i \hat{q}^k \hat{\tl{q}}^j \hat{\tl{q}}^l\right]  \left[ P_\pp^{ik}(\tb{k}) P_\pp^{jl}(\tb{k}) - \frac{1}{2} P_\pp^{ij}(\tb{k}) P_\pp^{kl}(\tb{k})\right] \\
= & \,\, G_\pp(q)G_\pp(\tl{q})  \left[\delta^{ik} \delta^{jl} - \delta^{ik} \hat{\tl{q}}^j \hat{\tl{q}}^l - \delta^{jl} \hat{q}^i \hat{q}^k +  \hat{q}^i \hat{q}^k \hat{\tl{q}}^j \hat{\tl{q}}^l \right] \\
&\times \left[\delta^{ik}\delta^{jl}-\delta^{ik}\hat{k}^j \hat{k}^l -\delta^{jl} \hat{k}^i \hat{k}^k + \frac{\hat{k}^i \hat{k}^k \hat{k}^j \hat{k}^l - \delta^{ij}\delta^{kl} + \delta^{ij}\hat{k}^k \hat{k}^l + \delta^{kl}\hat{k}^i \hat{k}^j}{2} \right] \\
& + G_\pp(q) G_\pl(\tl{q}) \left[\delta^{ik}\hat{\tl{q}}^j \hat{\tl{q}}^l - \hat{q}^i \hat{q}^k \hat{\tl{q}}^j \hat{\tl{q}}^l\right] \left[,,\right] \\
& + G_\pl(q) G_\pp(\tl{q}) \left[\delta^{jl} \hat{q}^i \hat{q}^k - \hat{q}^i \hat{q}^k \hat{\tl{q}}^j \hat{\tl{q}}^l \right] \left[,,\right] \\
& +  G_\pl(q) G_\pl(\tl{q}) \left[\hat{q}^i \hat{q}^k \hat{\tl{q}}^j \hat{\tl{q}}^l\right] \left[\delta^{ik}\delta^{jl}-\delta^{ik}\hat{k}^j \hat{k}^l -\delta^{jl} \hat{k}^i \hat{k}^k + \frac{\hat{k}^i \hat{k}^k \hat{k}^j \hat{k}^l - \delta^{ij}\delta^{kl} + \delta^{ij}\hat{k}^k \hat{k}^l + \delta^{kl}\hat{k}^i \hat{k}^j}{2} \right]  \\
= & \,\, G_\pp(q)G_\pp(\tl{q}) \left[-1 + \frac{1}{2}(\hat{\tl{q}}.\hat{k})^2 + \frac{1}{2}(\hat{q}.\hat{k})^2 + (\hat{q}.\hat{k})(\hat{\tl{q}}.\hat{k})-\frac{(\hat{q}.\hat{\tl{q}})^2}{2}+\frac{(\hat{q}.\hat{k})^2(\hat{\tl{q}}.\hat{k})^2}{2} \right]  \\
& + G_\pp(q) G_\pl(\tl{q}) \left[\frac{1}{2}-\frac{1}{2}(\hat{\tl{q}}.\hat{k})^2 + (\hat{q}.\hat{k})^2  -  (\hat{q}.\hat{k})(\hat{\tl{q}}.\hat{k}) +\frac{(\hat{q}.\hat{\tl{q}})^2}{2} - \frac{(\hat{q}.\hat{k})^2(\hat{\tl{q}}.\hat{k})^2}{2} \right] \\
&+ G_\pl(q) G_\pp(\tl{q})\left[\frac{1}{2}- \frac{1}{2}(\hat{q}.\hat{k})^2 + (\hat{\tl{q}}.\hat{k})^2 - (\hat{q}.\hat{k})(\hat{\tl{q}}.\hat{k}) +\frac{(\hat{q}.\hat{\tl{q}})^2}{2} - \frac{(\hat{q}.\hat{k})^2(\hat{\tl{q}}.\hat{k})^2}{2} \right]\\
& +  G_\pl(q) G_\pl(\tl{q}) \left[1- (\hat{\tl{q}}.\hat{k})^2-(\hat{q}.\hat{k})^2 + (\hat{q}.\hat{k})(\hat{\tl{q}}.\hat{k})-\frac{(\hat{q}.\hat{\tl{q}})^2}{2}+\frac{(\hat{q}.\hat{k})^2(\hat{\tl{q}}.\hat{k})^2}{2} \right]\\
\end{align*}
Similarly:
 
\begin{equation*}
    \Pi_{[2]} = \left[M^{il}(q) M^{jk}(\tl{q})\right] \Lambda^{ijkl}(\tb{k})
\end{equation*}
 
Evaluating the product and simplifying:
 
\begin{align*}
\Pi_{[2]} = & \,\, G_\pp(q)G_\pp(\tl{q}) \left[2 + \frac{1}{2}(\hat{\tl{q}}.\hat{k})^2 + \frac{1}{2}(\hat{q}.\hat{k})^2 - (\hat{q}.\hat{k})(\hat{\tl{q}}.\hat{k})+\frac{(\hat{q}.\hat{\tl{q}})^2}{2}+\frac{(\hat{q}.\hat{k})^2(\hat{\tl{q}}.\hat{k})^2}{2} \right]  \\
& + G_\pp(q) G_\pl(\tl{q}) \left[\frac{1}{2}-\frac{1}{2}(\hat{\tl{q}}.\hat{k})^2 +  (\hat{q}.\hat{k})(\hat{\tl{q}}.\hat{k}) -\frac{(\hat{q}.\hat{\tl{q}})^2}{2} - \frac{(\hat{q}.\hat{k})^2(\hat{\tl{q}}.\hat{k})^2}{2} \right] \\
&+ G_\pl(q) G_\pp(\tl{q})\left[\frac{1}{2}- \frac{1}{2}(\hat{q}.\hat{k})^2 + (\hat{q}.\hat{k})(\hat{\tl{q}}.\hat{k}) -\frac{(\hat{q}.\hat{\tl{q}})^2}{2} - \frac{(\hat{q}.\hat{k})^2(\hat{\tl{q}}.\hat{k})^2}{2} \right]\\
& +  G_\pl(q) G_\pl(\tl{q}) \left[\frac{(\hat{q}.\hat{\tl{q}})^2}{2} - (\hat{q}.\hat{k})(\hat{\tl{q}}.\hat{k})+\frac{(\hat{q}.\hat{k})^2(\hat{\tl{q}}.\hat{k})^2}{2} \right]\\
\end{align*}
Hence, $\Pi (\tb{k}) \equiv \Pi_{[1]}+\Pi_{[2]}$ becomes:
 
\begin{align*}
\Pi (\tb{k}) = & \,\, G_\pp(q)G_\pp(\tl{q}) \left[-1 + \frac{1}{2}(\hat{\tl{q}}.\hat{k})^2 + \frac{1}{2}(\hat{q}.\hat{k})^2 + (\hat{q}.\hat{k})(\hat{\tl{q}}.\hat{k})-\frac{(\hat{q}.\hat{\tl{q}})^2}{2}+\frac{(\hat{q}.\hat{k})^2(\hat{\tl{q}}.\hat{k})^2}{2} \right]  \\
&+ G_\pp(q) G_\pl(\tl{q}) \left[\frac{1}{2}-\frac{1}{2}(\hat{\tl{q}}.\hat{k})^2 + (\hat{q}.\hat{k})^2  -  (\hat{q}.\hat{k})(\hat{\tl{q}}.\hat{k}) +\frac{(\hat{q}.\hat{\tl{q}})^2}{2} - \frac{(\hat{q}.\hat{k})^2(\hat{\tl{q}}.\hat{k})^2}{2} \right] + G_\pl(q) G_\pp(\tl{q})\\
&\times\left[\frac{1}{2}- \frac{1}{2}(\hat{q}.\hat{k})^2 + (\hat{\tl{q}}.\hat{k})^2 - (\hat{q}.\hat{k})(\hat{\tl{q}}.\hat{k}) +\frac{(\hat{q}.\hat{\tl{q}})^2}{2} - \frac{(\hat{q}.\hat{k})^2(\hat{\tl{q}}.\hat{k})^2}{2} \right] +  G_\pl(q) G_\pl(\tl{q})\\
& \times\left[1- (\hat{\tl{q}}.\hat{k})^2-(\hat{q}.\hat{k})^2 + (\hat{q}.\hat{k})(\hat{\tl{q}}.\hat{k})-\frac{(\hat{q}.\hat{\tl{q}})^2}{2}+\frac{(\hat{q}.\hat{k})^2(\hat{\tl{q}}.\hat{k})^2}{2} \right] +  G_\pp(q)G_\pp(\tl{q})\\
&\times \left[2 + \frac{1}{2}(\hat{\tl{q}}.\hat{k})^2 + \frac{1}{2}(\hat{q}.\hat{k})^2 - (\hat{q}.\hat{k})(\hat{\tl{q}}.\hat{k})+\frac{(\hat{q}.\hat{\tl{q}})^2}{2}+\frac{(\hat{q}.\hat{k})^2(\hat{\tl{q}}.\hat{k})^2}{2} \right]  + G_\pp(q) G_\pl(\tl{q})\\
&\times\left[\frac{1}{2}-\frac{1}{2}(\hat{\tl{q}}.\hat{k})^2 +  (\hat{q}.\hat{k})(\hat{\tl{q}}.\hat{k}) -\frac{(\hat{q}.\hat{\tl{q}})^2}{2} - \frac{(\hat{q}.\hat{k})^2(\hat{\tl{q}}.\hat{k})^2}{2} \right] + G_\pl(q) G_\pp(\tl{q}) \left[\frac{1}{2}- \frac{1}{2}(\hat{q}.\hat{k})^2 +...\right] \\
&\left[... (\hat{q}.\hat{k})(\hat{\tl{q}}.\hat{k}) -\frac{(\hat{q}.\hat{\tl{q}})^2}{2} - \frac{(\hat{q}.\hat{k})^2(\hat{\tl{q}}.\hat{k})^2}{2} \right] +  G_\pl(q) G_\pl(\tl{q}) \left[\frac{(\hat{q}.\hat{\tl{q}})^2}{2} - (\hat{q}.\hat{k})(\hat{\tl{q}}.\hat{k})+\frac{(\hat{q}.\hat{k})^2(\hat{\tl{q}}.\hat{k})^2}{2} \right]\\
\end{align*}

\begin{multline}
\Rightarrow G_\pp(q)G_\pp(\tl{q}) \left[1 + (\hat{q}.\hat{k})^2 + (\hat{\tl{q}}.\hat{k})^2 +(\hat{q}.\hat{k})^2 (\hat{\tl{q}}.\hat{k})^2 \right]  \\
+ G_\pp(q) G_\pl(\tl{q}) \left[1 + (\hat{q}.\hat{k})^2-(\hat{\tl{q}}.\hat{k})^2 - (\hat{q}.\hat{k})^2 (\hat{\tl{q}}.\hat{k})^2 \right] \\
+ G_\pl(q) G_\pp(\tl{q})\left[1- (\hat{q}.\hat{k})^2 + (\hat{\tl{q}}.\hat{k})^2 - (\hat{q}.\hat{k})^2 (\hat{\tl{q}}.\hat{k})^2 \right]\\
+  G_\pl(q) G_\pl(\tl{q}) \left[1- (\hat{q}.\hat{k})^2 - (\hat{\tl{q}}.\hat{k})^2 + (\hat{q}.\hat{k})^2 (\hat{\tl{q}}.\hat{k})^2 \right] = \Pi (\tb{k}) \\
\end{multline}

Substituting equations for $\mu$ and $\tilde{\mu}$ in the above equation:
 
\begin{align*}
\Rightarrow\Pi (\tb{k}) = & \,\, G_\pp(q)G_\pp(\tl{q}) \left[1 +\mu^2 + \tl{\mu}^2 +\mu^2 \tl{\mu}^2 \right]  \\
&+ G_\pp(q) G_\pl(\tl{q}) \left[1 + \mu^2 - \tl{\mu}^2 - \mu^2 \tl{\mu}^2 \right] \\
&+ G_\pl(q) G_\pp(\tl{q})\left[1- \mu^2 + \tl{\mu}^2 - \mu^2 \tl{\mu}^2 \right]\\
&+  G_\pl(q) G_\pl(\tl{q}) \left[1- \mu^2 - \tl{\mu}^2 + \mu^2 \tl{\mu}^2 \right] \\
\Rightarrow\Pi (\tb{k}) = & \,\, G_\pp(q)G_\pp(\tl{q}) \left[1\times (1 +\mu^2) + \tl{\mu}^2 \times (1 + \mu^2) \right] \\
&+ G_\pp(q) G_\pl(\tl{q}) \left[1\times (1 + \mu^2) - \tl{\mu}^2 \times (1 + \mu^2) \right] \\
&+ G_\pl(q) G_\pp(\tl{q}) \left[1\times (1 - \mu^2) + \tl{\mu}^2 \times (1 - \mu^2) \right]\\
&+  G_\pl(q) G_\pl(\tl{q}) \left[1\times (1- \mu^2) - \tl{\mu}^2 \times (1 - \mu^2) \right] \\
\end{align*}
\begin{multline}
\Rightarrow\Pi (\tb{k}) = G_\pp(q)G_\pp(\tl{q}) \left[(1 + \mu^2) (1 + \tl{\mu}^2) \right] \\
+ G_\pp(q) G_\pl(\tl{q}) \left[(1 + \mu^2) (1 - \tl{\mu}^2) \right] \\
+ G_\pl(q) G_\pp(\tl{q}) \left[(1 - \mu^2) (1 + \tl{\mu}^2) \right] \\
+ G_\pl(q) G_\pl(\tl{q}) \left[(1- \mu^2) (1 - \tl{\mu}^2) \right] \\
\end{multline}
Or:
\[
\boxed{\Pi(k) = \Pi_{\pp\perp}(k) + \Pi_{\pp\pl}(k) + \Pi_{\pl\pp}(k) + \Pi_{\pl\pl}(k)}
\]
where:
$$ \Pi_{\pp\pp} (k) = \bar{w}^2  \int  \frac{d^3q}{(2\pi)^3} \left(1+\mu^2 \right)\left(1+\tl{\mu}^2 \right) G_\pp(q) G_\pp(\tl{q}) $$
$$ \Pi_{\pp\pl} (k) = \bar{w}^2  \int  \frac{d^3q}{(2\pi)^3} \left(1+\mu^2 \right)\left(1-\tl{\mu}^2 \right) G_\pp(q) G_\pl(\tl{q}) $$
$$ \Pi_{\pl\pp} (k) = \bar{w}^2  \int  \frac{d^3q}{(2\pi)^3} \left(1-\mu^2 \right)\left(1+\tl{\mu}^2 \right) G_\pl(q) G_\pp(\tl{q}) $$
$$ \Pi_{\pl\pl} (k) = \bar{w}^2  \int  \frac{d^3q}{(2\pi)^3} \left(1-\mu^2 \right)\left(1-\tl{\mu}^2 \right) G_\pl(q) G_\pl(\tl{q}) $$
\begin{figure}
\centering
\includegraphics[width=10cm]{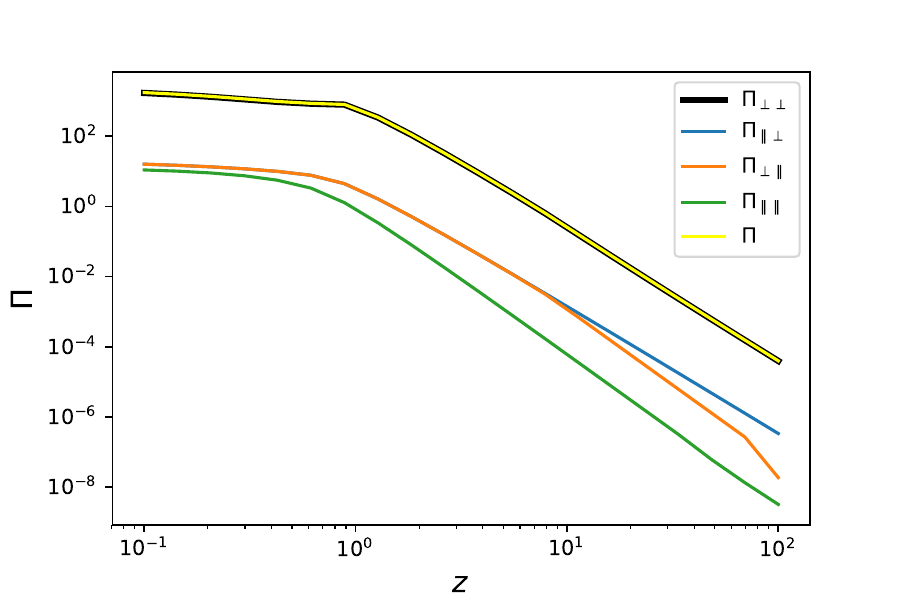}
\caption{\label{Pi variation}Variation of different combinations of $\Pi$ with transverse and longitudinal components and their sum vs the non dimensionalized wavenumber, $z$}
\end{figure}
 
Figure \ref{Pi variation} shows a plot of these tensors. The sound shell model is identified to be based on the $\Pi_{\parallel\parallel}$ component, while the turbulence is the $\Pi_{\perp\perp}$ component which is also equal to the sum of all components. Therefore the anisotropic stress tensor for the turbulent model is:
\beq
\label{Pipp}
 \Pi_{\pp\pp} (k) = \bar{w}^2  \int  \frac{d^3q}{(2\pi)^3} \left(1+\mu^2 \right)\left(1+\tl{\mu}^2 \right) G_\pp(q) G_\pp(\tl{q})
\eeq

\section{Gravitational wave relics}
The gravitational waves from the turbulent source should still be propagating through the universe. The total energy density varies with $a$, the scale factor, as $a^{-4}$ and the amplitude varies as $a^{-1}$ .
Also, since $\rho_c = 3H_0^2/8\pi G$, it is found that $\Omega \equiv \rho_{GW}/\rho_c$.
Changing from angular to linear frequency, $f=\om/2\pi$, and identifying $h_0$ as the current Hubble parameter $H_0$ in units of km/s Mpc$^{-1}$, the characteristic strain amplitude is given by:
\beq
h_c(f) = 1.263\times 10^{-18} \left(\frac{1Hz}{f}\right) [h_0^2 \Omega_G(f)]^{1/2}
\eeq
From this quantity at the epoch of turbulence, given by a scale factor $a_*$, the amplitude and frequency are reduced by a factor of $a_*/a_0$ with the temperature $T_*$ and the effective number of relativistic degrees of freedom $g_*$.

\subsubsection{Asymptotic limits}
When the low frequency regime (limit $\om \rightarrow 0$) is applied, the behaviour of the spectrum becomes independent of turbulence. The time delayed fourth-order correlation tensor $R\gg 1$ for fully developed turbulence with a Mach number $M$, defined as $M^2 \equiv \bar{U}^2/c_s^2$ ($\bar{U}: \text{root mean square velocity; }c_s: \text{speed of sound}$), gives a variance:
\beq
\Rightarrow h_c(f) \propto f^{-13/4}
\eeq
This modified turbulence spectrum has a radiation similar Kolmogorov turbulence spectrum in terms of the properties of the ability to be detected. The peak frequency depends on the eddy turnover time where the energy density peaks on the stirring scale, which is also independent of the turbulent cascade.

\section{Reproduced results}
In order to reproduce the results of this model, there are a couple more quantities to be defined. For two given times, $t_1$ and $t_2$, the time difference $t_-$ is defined as:
\beq
t_- = t_2 - t_1
\eeq
Given the following locally defined parameters for quick computations, the two important quantities needed for the further steps can be derived. Let:
\begin{align*}
    \zeta &= \frac{t_-}{\tau} = \frac{t_- \bar{U}}{\xi}\\
    z_u &= \frac{z}{\bar{U}}\\
     \beta &= iz_u \\
     \al &= \frac{\kappa_0^{2/3}}{12 C_K} \left(x^{4/3} + \tl{x}^{4/3} \right)
\end{align*}

\noindent \textbf{Rate of spectral density:}

\begin{align*}
\Delta_{\dot{h}} &= \int_{-\infty}^{\infty} dt_- \frac{\cos kt_-}{2} D(q,t_-) D(\tl{q},t_-)\\
&= \frac{1}{2} \int_{-\infty}^{\infty} d\zeta \frac{\xi}{\bar{U}} \cos\left(\frac{z\zeta}{\bar{U}}\right) \exp \left[-\frac{\pi}{4} \eta_q^2 t_-^2 \right] \exp\left[-\frac{\pi}{4} \eta_{\tl{q}}^2 t_-^2 \right]\\
&= \frac{1}{2}\frac{\xi}{\bar{U}}\sqrt{\frac{24\pi}{\left[\kappa_0^{2/3} \left(x^{4/3} + \tl{x}^{4/3} \right) \right]}} \exp{\left[-\frac{6C_K}{\kappa_0^{2/3}\left(x^{4/3} + \tl{x}^{4/3} \right) }\frac{z^2}{\bar{U}^2} \right]} \\
\end{align*}
\newline
\textbf{Gravitational wave power spectrum for stationary turbulence}
 
\begin{align*}
    \frac{d\tl{\mathcal{P}}_{\dot{h}}}{dt} &= \frac{1}{2}\frac{\xi}{\bar{U}} \sqrt{\frac{24\pi}{\left[\kappa_0^{2/3} \left(x^{4/3} + \tl{x}^{4/3} \right) \right]}} \exp{\left[-\frac{6C_K}{\kappa_0^{2/3}\left(x^{4/3} + \tl{x}^{4/3}\right)}\frac{z^2}{\bar{U}^2} \right]}\\
    \mathcal{P}_{\dot{h}} &= \left(16 \pi G \right)^2 \frac{\bar{w}^2}{4\pi c_s} \tau_T \times \frac{d\mathcal{P}_{\dot{h}}}{dt} \\
    \rho_{GW} &= \left(16 \pi G \right)^2 \frac{\bar{w}^2}{4\pi c_s} \frac{k^3}{2\pi^2} \frac{\tau_T}{32\pi G} \frac{d\mathcal{P}_{\dot{h}}}{dt} \quad ; \quad \Om_{GW} = \left(16 \pi G \right)^2 \frac{\bar{w}^2}{4\pi c_s} \frac{k^3}{2\pi^2} \frac{\tau_T}{32\pi G \rho_c} \frac{d\mathcal{P}_{\dot{h}}}{dt}
\end{align*}

Since $G\rho_c = 3H^2/8\pi $:
 
\begin{align*}
\Omega_{GW} &= \left(16 \pi G \right)^2 \frac{\bar{w}^2}{4\pi c_s} \frac{k^3}{2\pi^2} \frac{\tau_T}{12 H^2} \frac{d\mathcal{P}_{\dot{h}}}{dt} \\
&= \frac{2}{3\pi^3} \left(\frac{16\pi^2}{45} \kappa_0^{4/3} \right) z^3 (H\tau_T) \left(\frac{H\xi}{c_s}\right) \bar{U}^3 \frac{d{\tl{\mathcal{P}}_{\dot{h}}}^\prime}{dt}\\
\Rightarrow \frac{\Om_{GW}}{(H\tau_T) (H\xi) \bar{U}^3} &= \frac{32\kappa_0^{4/3}}{135\pi c_s} \cdot z^3 \cdot \frac{d{\tl{\mathcal{P}}_{\dot{h}}}^\prime}{dt}\\
\end{align*}
 
\begin{figure}
\centering
\includegraphics[width=10cm]{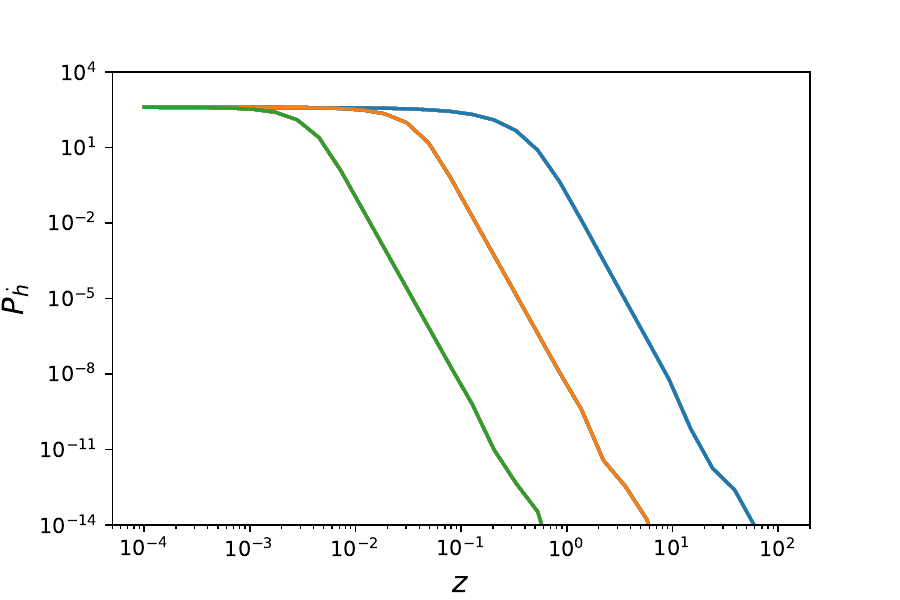}
\caption{\label{SdrGKK}Variation of rate of spectral density (model 1) against non dimensionalized wavenumber}
\end{figure}

\begin{figure}
\centering
\includegraphics[width=10cm]{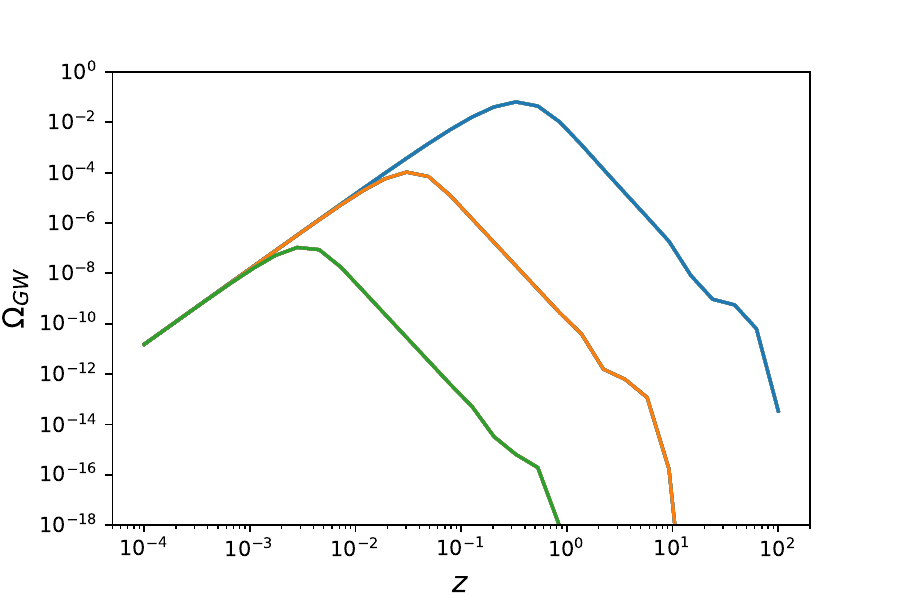}
\caption{\label{OGKK}Variation of the gravitational wave power spectrum (model 1) against non dimensionalized wavenumber}
\end{figure}

\begin{figure}
\centering
\includegraphics[width=10cm]{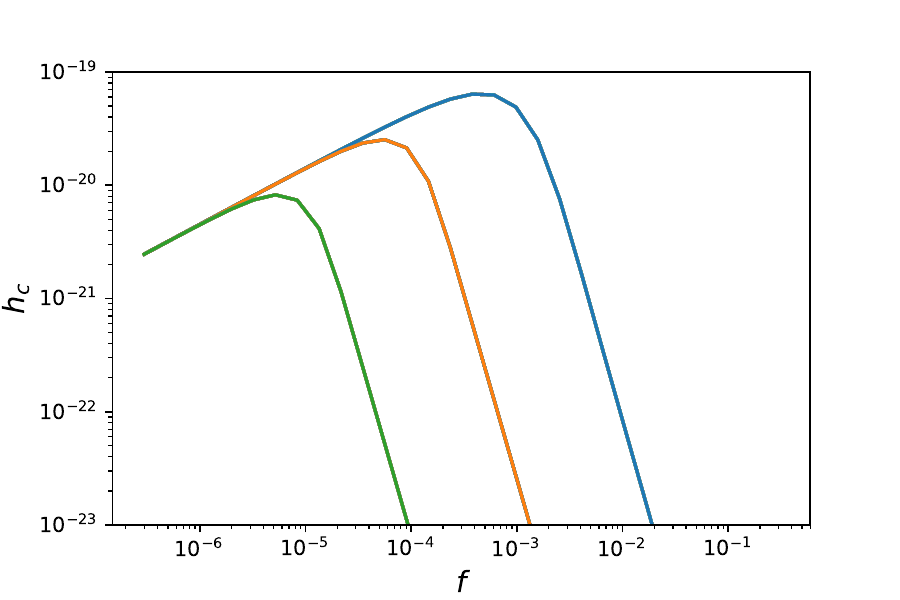}
\caption{\label{hGKK}Variation of gravitational wave amplitude with frequency (model 1) for Mach numbers, $M = 1$ (blue), $M = 0.1$ (red) and $M = 0.01$ (green)}
\end{figure}

Subsequently, the plots of various spectrum were obtained consistent with the stationary turbulence model \citep{Gogoberidze:2007an} as seen in figures \ref{SdrGKK}, \ref{OGKK} and \ref{hGKK}.

\chapter{Top hat correlation model}
\label{CDS}
The importance of relics is that they give information about the hot, dense universe that existed billions of years ago. Gravitational wave relics are of high interest because of their negligible interaction with matter and radiation, thus, facilitating the direct observation of the early universe. As mentioned earlier, the source of these relic gravitational waves are taken to be the broken phase bubbles that underwent first order phase transition. The collision of bubbles in extreme physical conditions, particularly Reynolds number, not only instigates energy to be injected in the fluid until the end of phase transition, but also rapidly develops a turbulence in the fluid, which is assumed to have lasted until the bubbles had percolated. But the dissipation of the turbulence does not cease until many Hubble times. This, along with the properties of the source, enables the determination of the nature of the universe before inflation.

\section{Short lasting source}
For a fluid field of plasma at Reynolds number of the order $\sim 10^13$, turbulence arises from perturbations the fluid is subjected to: bubble collisions, that which can be referred to as the source of fluid stirring. In the period that this project is based on, the expansion of the universe is neglected by considering a finite, continuous, short-lasting source: the source was active only during the short interval of phase transition, i.e. the characteristic time over which the turbulence was on. The energy spectrum of such a source can be written as:
\beq
\frac{d\Om_{GW}}{d \log k} \propto \int dt_- \cos (kt_-) \tl{\Pi}(k, t_1, t_2)
\eeq
where, $t_- = t_2 - t_1$ and $\tl{\Pi}(k, t_1, t_2)$ is the anisotropic stress tensor given by the unequal time correlator of the source:
\beq
\langle \tl{\Pi}_{ij} (\mathbf{k}, t_1)  \tl{\Pi}_{ij}^* (\mathbf{q}, t_2) \rangle = (2\pi)^3 \delta(\mathbf{k} - \mathbf{q}) \tl{\Pi}(k, kt_1, kt_2)
\eeq
The time over which the cascade set in, with the plasma having a fluid velocity $v_f$ stirred on the characteristic length scale ($L$), is called as the eddy turnover time, given by $\tau_L = L/2v_f$. This implies that the source lasted only for one $\tau_L$.

\subsection{Power spectrum of velocity field for equal times}
The fluid flow has a constant density making the turbulent velocity field incompressible and divergence-free and indicating a projector, $P_{\perp\perp} \equiv \delta_{ij} - \hat{k}_i \hat{k}_j$. At equal times, its power spectrum, $G_{\perp\perp}(k,t)$ can be defined by:
\beq
\langle v_i(\mathbf{k}, t) v_j^* (\mathbf{q}, t) = (2\pi)^3 \delta (\mathbf{k}-\mathbf{q}) P_{\perp\perp} G_{\perp\perp}(k,t)
\eeq
For a dimensionless quantity, $K \equiv kL/2\pi$:
\beq
G_{\perp\perp} (K) \propto L^3 \frac{K^2}{(1+K^2)^{17/6}} \quad\quad\quad 0 \leq K \leq L/\lambda
\eeq
where, $\lambda$ is the Kolmogorov micro-scale below which the turbulence is off. When the phase transition is complete, the turbulence starts decaying rapidly and freely, dissipating the injected kinetic energy.

\subsection{Power spectrum of velocity fields for unequal times}
The velocity power spectrum for unequal times is modified as:
\beq
\langle v_i(\mathbf{k}, t_1) v_j^* (\mathbf{q}, t_2) = (2\pi)^3 \delta (\mathbf{k}-\mathbf{q}) P_{\perp\perp} G_{\perp\perp}(k,t_1, t_2)
\eeq
To achieve this, the equal time power spectrum is multiplied by a temporal decorrelation function defined over the eddy turnover time. Here, the Kraichnan exponential decorrelation which takes a Gaussian form is chosen for modeling:
$$\eta (t_1, t_2) = \exp \left[ -\frac{\pi}{4} \frac{(t_1 - t_2)^2}{\tau_L^2} \right]$$

\subsection{Power spectrum of anisotropic stress}
The anisotropic stress for unequal times defined above can be further written as:
\beq
\langle \tl{\Pi}_{ij} (\mathbf{k}_1, t_1)  \tl{\Pi}_{ij}^* (\mathbf{k}_1, t_2) \rangle = 
\mathcal{P}_{abcd} \int \frac{d^3 p}{(2\pi)^3} \int \frac{d^3 q}{(2\pi)^3} \langle v_a (\mathbf{k}_1 - \mathbf{p}, t_1) v_b (\mathbf{p}, t_1) v_c^* (\mathbf{k}_2 - \mathbf{q}, t_2) v_d^*(\mathbf{q}, t_2) \rangle
\eeq
where $\mathcal{P}_{abcd}$ can be computed from:
\beq
\mathcal{P}_{abcd} \equiv \left(P_{ma} P_{nb} - \frac{1}{2}P_{mn}P_{ab} \right) (\mathbf{k}) \left(P_{mc} P_{nd} - \frac{1}{2}P_{mn}P_{cd} \right) (\mathbf{q})
\eeq
The spectrum of the anisotropic stress, when integrated, should remain positive. To obtain this condition consistent with the Kraichnan decorrelation, introducing the top hat approximation:
\beq
    \Pi (k, t_1, t_2) \propto \Pi(k, t_1, t_1) \Theta (t_2 - t_1) \Theta \left(\frac{x_c}{k} - (t_2 - t_1) \right) \\
+ \Pi(k, t_2, t_2) \Theta (t_1 - t_2) \Theta \left(\frac{x_c}{k} - (t_1 - t_2) \right)
\eeq
where, $x_c = 1$, suggesting that $\Pi(k, t_1, t_2)$ is symmetric and is correlated if $kt_- < x_c$; else, uncorrelated.
Recalling the velocity power spectrum, for a dimensionless variable, $\bar{y} := (t_1 - t_{in})/\tau_L$, where $t_{in}$ is the time at which the phase transition starts, it is seen that:
\beq
\Pi_v(K, \bar{y}, \bar{y}) \propto L^3(\bar{y}) \mathcal{I}_v(K, \bar{y}, \bar{y})
\eeq
with $\mathcal{I}_v (K, \bar{y}, \bar{y})$ defined as:
\beq
\mathcal{I}_v (K, \bar{y}, \bar{y}) = \int_0^\infty dQ \frac{Q^4}{(1+Q^2)^{17/6}} \int_{-1}^1 d \chi (1+\chi^2) \frac{2K^2 + Q^2(1+\chi^2) - 4KQ\chi}{(1+K^2 - 2K Q\chi + Q^2)^{17/6}}
\eeq
Here, $Q \equiv Lq/2\pi$ and $\chi$ is equivalent to the factor $\mu$ defined earlier. This is the best approximation for the turbulent decorrelation that is being considered. Hence, this is chosen while formulating for the new model.

\section{Replicated results}
To obtain the plots, the rate of spectral density needs to be determined. Recalling:
$$\zeta = \frac{t_-}{\tau} = \frac{t_- \bar{U}}{\xi}\quad ; \quad z_u = \frac{z}{\bar{U}} \quad ; \quad \beta = iz_u $$
And redefining to simplify integration:
$$\al = \frac{x^2 + \tl{x}^2}{2\pi {\bar{U}^2}} $$
The following is obtained:
\begin{align*}
\Delta_{\dot{h}} &= \int_{-\infty}^{\infty} dt_- \frac{\cos kt_-}{2} D(q,t_-) D(\tl{q},t_-)\\
&= \frac{1}{2} \int_{-\infty}^{\infty} d\zeta \frac{\xi}{\bar{U}} \cos\left(\frac{z\zeta}{\bar{U}}\right) \exp \left[-\frac{1}{4\pi}  q^2 t_-^2 \right] \exp\left[-\frac{1}{4\pi} {\tl{q}}^2 t_-^2 \right]\\
&= \frac{1}{2}\frac{\xi}{\bar{U}} \int_{-\infty}^{\infty} d\zeta \left(\frac{e^{iz_u\zeta}+ e^{-iz_u\zeta}}{2}\right)\exp \left[-\frac{1}{4\pi {\bar{U}^2}} \left(x^2 + \tl{x}^2 \right) \zeta^2 \right] \\
&= \frac{1}{2}\frac{\xi}{\bar{U}}\sqrt{\frac{4\pi^2 {\bar{U}^2}}{x^2 + \tl{x}^2 }} \exp{\left[-\frac{\pi z^2}{x^2 + \tl{x}^2}  \right]} \\
\end{align*}
 
\begin{figure}
\centering
\includegraphics[width=10cm]{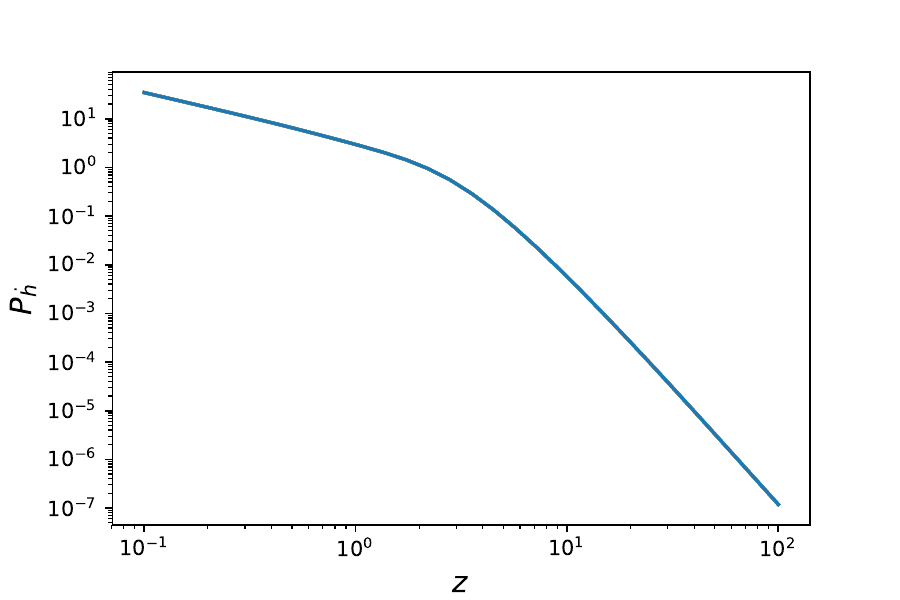}
\caption{\label{SdrCDS} Variation of rate of spectral density (model 2) against non dimensionalized wavenumber}
\end{figure}

\begin{figure}
\centering
\includegraphics[width=10cm]{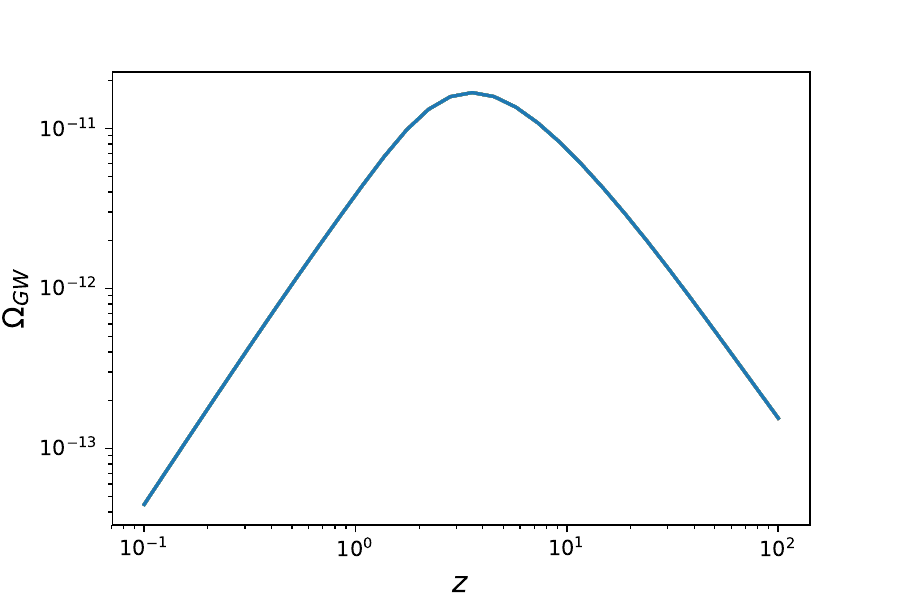}
\caption{\label{OCDS} Variation of the power spectrum (model 2) against non dimensionalized wavenumber}
\end{figure}

\begin{figure}
\centering
\includegraphics[width=10cm]{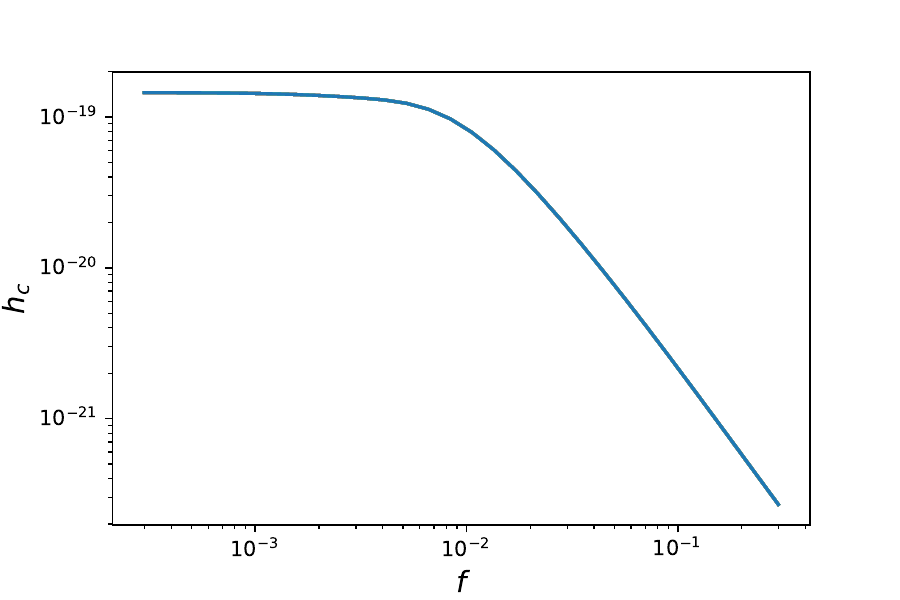}
\caption{\label{hCDS} Variation of amplitude with frequency (model 2) for Mach, $M = 1$}
\end{figure}

These results are targeted to mimic the results from the \citep{Caprini:2009yp} second model; the graphs can be seen from the figures \ref{SdrCDS}, \ref{OCDS} and \ref{hCDS}.

\chapter{The sweeping decorrelation model}
\begin{figure}
\centering
\includegraphics[width=10cm]{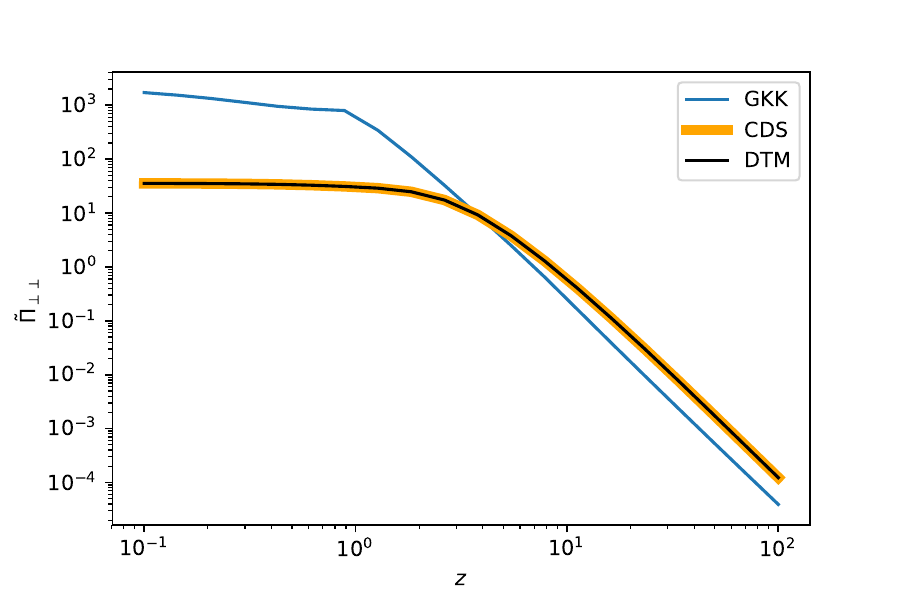}
\caption{\label{etc} Variation of the anisotropic stress of all models against non dimensionalized wavenumber}
\end{figure}

Although the Kraichnan formulation used up to this point for establishing the correlations in the Eulerian spacetime is satisfactory for turbulence, it fails to concur for fluids at high Reynolds number. Here, the 'sweeping hypothesis' given by Kraichnan is adopted to resolve this problem. It assumes that the spacetime correlations are determined predominantly by the root mean square velocity of the plasma $\bar{U}^2$ called the sweeping velocity. This changes the decorrelation function from \ref{gkkexp} to:
\beq
f(\eta_k, t_-) = \exp{\left(-\frac{1}{2}\bar{U}^2 \eta_k^2 t_-^2 \right)}
\eeq
This is utilized in the top hat correlation model for normalizing the velocity power spectrum, determining the anisotropic stress tensor given by \ref{Pipp} and for further calculations just as in the previous cases. The equal time correlators of the two reference models  and the new model is compared in figure \ref{etc}. When converted to unequal times, it is interesting to take a look at the contour maps traced out by $\Pi$ in the appendix \ref{contours}.
 
The next step is to derive the expression for its rate of spectral density - with the unequal time correlators - with which the three spectral plots can be obtained. Again, locally defining the following parameters to ease calculations:
$$\al = x^2 + \tl{x}^2 $$
\begin{align*}
    \begin{split}
        \Delta_{\dot{h}} &= \int_{-\infty}^{\infty} dt_- \frac{\cos kt_-}{2} D(q,t_-) D(\tl{q},t_-)\\
        &= \frac{1}{2} \int_{-\infty}^{\infty} d\zeta \frac{\xi}{\bar{U}} \cos\left(\frac{z\zeta}{\bar{U}}\right) \exp \left[-\frac{1}{2} {\bar{U}}^2 q^2 t_-^2 \right] \exp\left[-\frac{1}{2} {\bar{U}}^2 {\tl{q}}^2 t_-^2 \right]\\
        &= \frac{1}{2}\frac{\xi}{\bar{U}}\sqrt{\frac{2\pi }{x^2 + \tl{x}^2 }} \exp{\left[-\frac{1}{2} \left(x^2 + \tl{x}^2\right)^{-1} \frac{z^2}{\bar{U}^2} \right]} \\
    \end{split}
\end{align*}
 
\begin{figure}
\centering
\includegraphics[width=10cm]{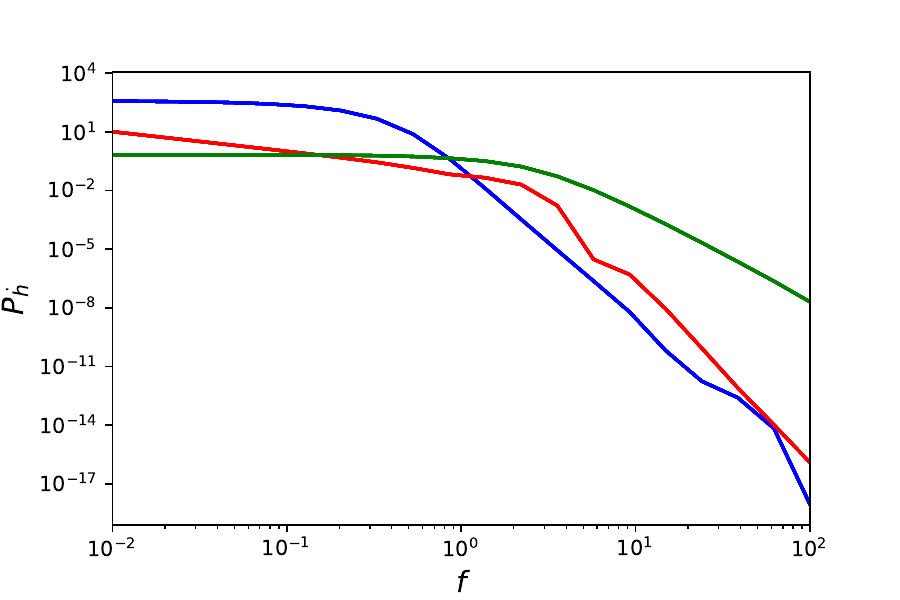}
\caption{\label{allSdr} Variation of rate of spectral density (all three models) against non dimensionalized wavenumber}
\end{figure}

\begin{figure}
\centering
\includegraphics[width=10cm]{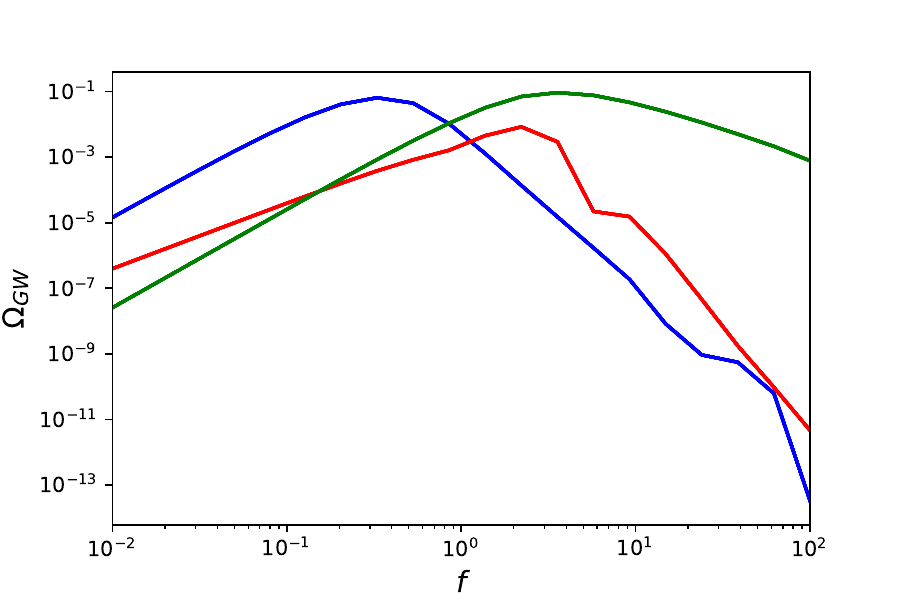}
\caption{\label{allO} Variation of gravitational wave power spectrum (all three models) against non dimensionalized wavenumber}
\end{figure}

\begin{figure}
\centering
\includegraphics[width=10cm]{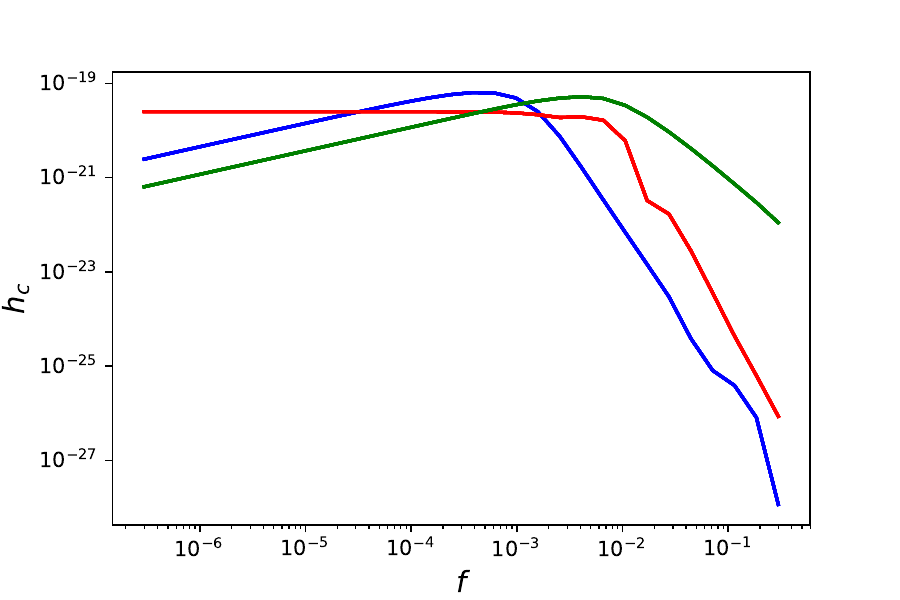}
\caption{\label{allh} Variation of amplitude with frequency (all three models) for Mach, $M=1$}
\end{figure}

Figures \ref{allSdr}, \ref{allO} and \ref{allh} shows the three required spectra of the new model (colour green) compared against the reference models  from chapter \ref{GKK} (colour blue) and \ref{CDS} (colour red).

\section{Analysis of the results}
\begin{itemize}
    \item \textbf{{Shear stress correlator:}}
    Shear stress correlator was seen to have a power law variation of $k^{-17/6}$ in all cases before integration and $k^{-11/3}$ after integration.
    \item \textbf{{Rate of spectral density:}}
    The rate of spectral density, which is the time derivative ($t_+ \equiv t_1 + t_2$) of the spectral density, starts varying as it approaches unity with a power law $k^{-13/3}$ as expected.
    \item \textbf{{Power spectrum:}}
    The power spectrum of the new model shows a $k^3$ variation until it peaks at $\sim 0.093$ for $z \approx 3.56$ and starts decaying with a $k^{-5/3}$ variation.
    \item \textbf{{Amplitude versus frequency:}}
    The amplitude of the sweeping decorrelation varies as $\sqrt{k}$ peaking at $\sim 8.3 \times 10^{-21}$ for $z \approx 1.37$ and starts dropping as a  $k^{-11/6}$ variation.
\end{itemize}
These results prominently demonstrate the effect of combining the vital terms of the two reference models  by retaining the behaviours of both sets of variations. The barrier of being restricted at high Reynolds number is, thus, broken by this model. Therefore, it is inferred that this model is better than the reference models.

\chapter{Conclusion}
\label{chap:conc}
The results that were emulated from the reference models had certain limitations. The new model has not only accounted for these limitations by applying some corrections to the decorrelation function, but also has retained the crucial characteristics of and from both models.

\section{Reference models  versus the new model}
The basic idea of this project was to determine the velocity spectrum, power spectrum, amplitude versus frequency spectrum, and their peaks. Since the reproduced results were made to match the original graphs, the same sequence of procedures were incorporated to obtain the new behaviours. The final plots of the procured model, produced an analytical variation that retained the range of the spectra, in spite of the corrections that were made. Also, the slopes of the spectra obeyed the stationary turbulence model in the lower end of $z$ and obeyed a modified power law for higher $z$, i.e. past the peaks.

\subsection{Limitations of the reference models }
The stationary turbulence model adopted the temporal decorrelation given by the Kraichnan exponential function. The approximations made by this function is limited \footnote{\url{https://ac.els-cdn.com/S0022247X98961911/1-s2.0-S0022247X98961911-main.pdf?_tid=68c33dbf-5a5f-4875-92e9-cd5fa241806c&acdnat=1534427303_8ed0fa74b794ab266a72f7b205a12443}} to low Reynolds number. Also, this model considers the traceless part of the energy-momentum tensor alone.
 
The authors' choice of the Heaviside function for the top hat correlation function has directly accounted for the power spectrum without any need for the velocity power spectrum.

\subsection{Scope of the proposed model}
The new model, as expected, addresses the above limitations by:
\begin{enumerate}
    \item Implementing 'sweeping' and enhancing the threshold of attainable Reynolds number
    \item Preserving the behaviour of the top hat correlator by postulating an alternate form for the Heaviside function so as to acquire the de-coherence function.
\end{enumerate}
The limitation of this model is its inability to account for the freely decaying turbulence. This requires an inclusion of variable root mean square factor in the spectral equations.
 
Another limitation is the adaptability to the turbulence being on or off by recognizing the status of the source. This is because of the assumption that the turbulence is short lasting having a period fairly lesser than one Hubble time, in an attempt to ignore the expansion of the universe during the phase transitions, which seems quite far-fetched for a hot, dense, infant universe. This condition is already treated by \citep{Caprini:2009yp} the second model, but it transcends the scope of this project.

\section{Detection of gravitational waves}
Hydrodynamic turbulence driven by the first-order phase transitions have a convenient property of gravitational wave emission peaking at the Kolmogorov de-coherence frequency. The power law and the normalization scale of the gravitational wave background energy density spectrum plays a critical role here. Bubble expansion itself cannot generate gravitational waves until its symmetry is broken, causing a macroscopic motion of cosmic plasma due to expansion, creating turbulence which becomes a significant source of the stochastic gravitational wave background. Detection of such backgrounds are believed to help solve the Higgs particle and high-energy particle physics. It can also verify the general relativity in robust gravitational fields. Due to its minimal interaction with matter and radiation in the universe, the relic gravitational wave acts as the fingerprint of the early universe and can answer a lot of questions about the moment of creation. Future projects like Laser Interferometer Space Antenna (eLISA), Deci-Hertz Interferometer Gravitational wave Observatory (DECIGO) and Square Kilometre Array (SKA) work with the aim of improving the experimental sensitivities with such analyses to accomplish a successful detection in the near future. 

\chapter*{Acknowledgements}
\renewcommand{\baselinestretch}{\linespacing}
\small\normalsize
I express my heartfelt gratitude to my supervisor, Prof Mark Hindmarsh, who has helped in every step throughout my degree and offered me support in every way possible.
 
I would like to show my appreciation to my parents, brother and family who have been a moral support in this tough year.
 
I would like to convey my sincere thanks to my previous Astrophysics teacher, Dr. Bharat Sharma, who has thrived to make me comfortable with anything related to Physics, irrespective of the time, distance or effort.
 
Finally, I am very grateful to all my seniors and friends, who have directly or indirectly helped me for this project.

\clearpage
\phantomsection
\addcontentsline{toc}{chapter}{Bibliography}
\bibliography{bib}

@article{Hindmarsh:2019phv,
    author = "Hindmarsh, Mark and Hijazi, Mulham",
    title = "{Gravitational waves from first order cosmological phase transitions in the Sound Shell Model}",
    eprint = "1909.10040",
    archivePrefix = "arXiv",
    primaryClass = "astro-ph.CO",
    reportNumber = "NORDITA-2019-083, HIP-2019-29/TH",
    doi = "10.1088/1475-7516/2019/12/062",
    journal = "JCAP",
    volume = "12",
    pages = "062",
    year = "2019"
}

@article{Cutting:2020nla,
    author = "Cutting, Daniel and Escartin, Elba Granados and Hindmarsh, Mark and Weir, David J.",
    title = "{Gravitational waves from vacuum first order phase transitions II: from thin to thick walls}",
    eprint = "2005.13537",
    archivePrefix = "arXiv",
    primaryClass = "astro-ph.CO",
    reportNumber = "HIP-2020-13/TH",
    doi = "10.1103/PhysRevD.103.023531",
    journal = "Phys. Rev. D",
    volume = "103",
    number = "2",
    pages = "023531",
    year = "2021"
}

@article{Hindmarsh:2020hop,
    author = {Hindmarsh, Mark B. and L{\"u}ben, Marvin and Lumma, Johannes and Pauly, Martin},
    title = "{Phase transitions in the early universe}",
    eprint = "2008.09136",
    archivePrefix = "arXiv",
    primaryClass = "astro-ph.CO",
    reportNumber = "MPP-2020-163, HIP-2020-27/TH",
    doi = "10.21468/SciPostPhysLectNotes.24",
    journal = "SciPost Phys. Lect. Notes",
    volume = "24",
    pages = "1",
    year = "2021"
}

@article{Caprini:2015tfa,
      author         = "Caprini, Chiara",
      title          = "{Stochastic background of gravitational waves from
                        cosmological sources}",
      booktitle      = "{Proceedings, 10th International LISA Symposium:
                        Gainesville, Florida, USA, May 18-23, 2014}",
      journal        = "J. Phys. Conf. Ser.",
      volume         = "610",
      year           = "2015",
      number         = "1",
      pages          = "012004",
      doi            = "10.1088/1742-6596/610/1/012004",
      eprint         = "1501.01174",
      archivePrefix  = "arXiv",
      primaryClass   = "gr-qc",
      SLACcitation   = "%%CITATION = ARXIV:1501.01174;%%"
}

@article{Linde:1978px,
      author         = "Linde, Andrei D.",
      title          = "{Phase Transitions in Gauge Theories and Cosmology}",
      journal        = "Rept. Prog. Phys.",
      volume         = "42",
      year           = "1979",
      pages          = "389",
      doi            = "10.1088/0034-4885/42/3/001",
      reportNumber   = "LEBEDEV-78-166",
      SLACcitation   = "%%CITATION = RPPHA,42,389;%%"
}

@article{Gleiser:1998kk,
      author         = "Gleiser, Marcelo",
      title          = "{Phase transitions in the universe}",
      journal        = "Contemp. Phys.",
      volume         = "39",
      year           = "1998",
      pages          = "239-253",
      doi            = "10.1080/001075198181937",
      eprint         = "hep-ph/9803291",
      archivePrefix  = "arXiv",
      primaryClass   = "hep-ph",
      SLACcitation   = "%%CITATION = HEP-PH/9803291;%%"
}

@article{Espinosa:2010hh,
      author         = "Espinosa, Jose R. and Konstandin, Thomas and No, Jose M.
                        and Servant, Geraldine",
      title          = "{Energy Budget of Cosmological First-order Phase
                        Transitions}",
      journal        = "JCAP",
      volume         = "1006",
      year           = "2010",
      pages          = "028",
      doi            = "10.1088/1475-7516/2010/06/028",
      eprint         = "1004.4187",
      archivePrefix  = "arXiv",
      primaryClass   = "hep-ph",
      reportNumber   = "CERN-PH-TH-2010-027",
      SLACcitation   = "%%CITATION = ARXIV:1004.4187;%%"
}

@book{Rezzolla:2013dea,
    author = "Rezzolla, Luciano and Zanotti, Olindo",
    title = "{Relativistic Hydrodynamics}",
    doi = "10.1093/acprof:oso/9780198528906.001.0001",
    isbn = "978-0-19-174650-5, 978-0-19-852890-6",
    publisher = "Oxford University Press",
    month = "9",
    year = "2013"
}

@article{Gogoberidze:2007an,
      author         = "Gogoberidze, Grigol and Kahniashvili, Tina and Kosowsky,
                        Arthur",
      title          = "{The Spectrum of Gravitational Radiation from Primordial
                        Turbulence}",
      journal        = "Phys. Rev.",
      volume         = "D76",
      year           = "2007",
      pages          = "083002",
      doi            = "10.1103/PhysRevD.76.083002",
      eprint         = "0705.1733",
      archivePrefix  = "arXiv",
      primaryClass   = "astro-ph",
      SLACcitation   = "%%CITATION = ARXIV:0705.1733;%%"
}

@article{Caprini:2009yp,
      author         = "Caprini, Chiara and Durrer, Ruth and Servant, Geraldine",
      title          = "{The stochastic gravitational wave background from
                        turbulence and magnetic fields generated by a first-order
                        phase transition}",
      journal        = "JCAP",
      volume         = "0912",
      year           = "2009",
      pages          = "024",
      doi            = "10.1088/1475-7516/2009/12/024",
      eprint         = "0909.0622",
      archivePrefix  = "arXiv",
      primaryClass   = "astro-ph.CO",
      SLACcitation   = "%%CITATION = ARXIV:0909.0622;%%"
}

\chapter{Appendix: Algebra}
\label{apxa}
\section*{Determination of $\Pi$ in the sound shell model}
The parametric definitions in chapter \ref{GKK} and chapter \ref{CDS} are used here and the factor $\mu$ is assumed to be independant of $q$ for simplicity.
 
Recalling $\eta$ as:
$$\eta_q^2 = \frac{1}{2\pi} \ep^{2/3} q^{4/3} \quad ; \quad \eta_k^2 = \frac{1}{2\pi}\ep^{2/3} k^{4/3}$$
For:
$$E_q = C_q \ep^{2/3} q^{-5/3}$$
The quantity $\Pi$ is given by:
$$ \Pi (k, t_1, t_2) = 4\bar{w}^2  \int_{-1}^{1} d\mu \left(1-\mu^2 \right)^2 \int_{0}^{k} \frac{q^2 dq}{(2\pi)^2}\frac{q^2}{\tl{q}^2} P_v(q) P_v(\tl{q}) \exp{\left[-\frac{\pi}{4}(\eta_{q}^{2} + \eta_{\tl{q}}^{2}) t_{-}^2 \right]} $$
where $P_v(q)$ is defined as:
$$P_v(q) = 4\pi^2\frac{E_q}{q^2} = 4\pi^2 C_q\ep^{2/3}q^{-11/3}$$
Defining $\xi$, to non-dimensionalize the equations, such that:
$$\xi \propto \frac{1}{k_0}$$
When $q \ll k$, approximation follows as $\tl{q}\approx k$, giving:
 
\begin{align*}
    \Pi_A (k, t_1, t_2) = & \,\, 4\bar{w}^2 \frac{P_v(k)}{k^2}\exp {\left[-\frac{\pi}{4}\eta_{k}^{2} t_{-}^2 \right]} \int_{-1}^{1} d\mu  \left(1-\mu^2 \right)^2  \int_{0}^{k} \frac{q^2 dq}{(2\pi)^2}{q^2} P_v(q)\\
 = & \,\, 4\bar{w}^2 \frac{P_v(k)}{k^2}\exp {\left[-\frac{\pi}{4}\eta_{k}^{2} t_{-}^2 \right]} \int_{-1}^{1} d\mu  \left(1-\mu^2 \right)^2  \int_{0}^{k} \frac{{q^4}dq}{(2\pi)^2} \frac{4\pi^2 C_q\ep^{2/3}}{q^{11/3}}\exp{\left[-\frac{\pi}{4}\eta_{q}^{2} t_{-}^2 \right]} \\
= & \,\, 4\bar{w}^2 \frac{4\pi^2 C_k\ep^{4/3}k^{-11/3}}{k^2}e^{-\frac{\pi}{4}\eta_{k}^{2} t_{-}^2} \left[\mu - \frac{2\mu^3}{3} + \frac{\mu^5}{5}\right]_{-1}^1  \int_{0}^{k} d(q\xi)\frac{C_q}{\xi}\frac{(q\xi)^{1/3}}{2\pi\xi^{1/3}}e^{-\frac{\pi}{4}\eta_{q}^{2} t_{-}^2}
\end{align*}

$$\eta_q^2 \rightarrow \eta_{q\xi}^2 \Rightarrow \eta_{q\xi}^2 = \frac{\ep^{2/3}q^{4/3}}{2\pi} \frac{\xi^{4/3}}{\xi^{4/3}} = \frac{\ep^{2/3}}{2\pi} \frac{(q\xi)^{4/3}}{\xi^{4/3}}$$
And since $P_v(q<k_0) = 0$, the lower limit of the integral changes from $0$ to $k_0$.
 
Hence,
 
\begin{equation*}
\begin{split}
\Rightarrow \tl{\Pi}_A (k\xi, t_1, t_2) = & \,\, 4\bar{w}^2 \frac{4\pi^2 C_k\ep^{4/3}k^{-17/3}}{\xi^{4/3}(\xi^{-17/3}\xi^{17/3})}e^{-\frac{\pi}{4}\eta_{k}^{2} t_{-}^2} \frac{16}{15} \int_{k_0\xi}^{k\xi} d(q\xi)C_q(q\xi)^{\frac{1}{3}}\exp {\left[-\frac{\pi}{4}\left(\frac{\ep^{2/3}}{2\pi}\frac{(q\xi)^{4/3}}{\xi^{4/3}}\right) t_{-}^2 \right]}\\
  = & \,\, \frac{16\bar{w}^2}{15} \frac{16\pi^2 C_k^2\ep^{4/3}(k\xi)^{-17/3}}{\xi^{-13/3}e^{{\pi\eta_{k}^{2} t_{-}^2/4}}} \left[-\frac{3}{4}\left(\frac{8\xi^{4/3}}{\ep^{2/3}t_-^2}\right)\exp {\left\{-\frac{\pi}{4}\left(\frac{\ep^{2/3}}{2\pi}\frac{(q\xi)^{4/3}}{\xi^{4/3}}\right) t_{-}^2 \right\}}\right]_{k_0\xi}^{k\xi}\\
 = & \,\, \frac{4\bar{w}^2}{5}\frac{16\pi^2 C_k^2 \ep^{4/3}\xi^{13/3}}{(k\xi)^{17/3}}e^{-\frac{\pi}{4}\eta_{k}^{2} t_{-}^2}\left(\frac{8\xi^{4/3}}{\ep^{2/3}t_-^2}\right) \\
 & \times \left\{\exp {\left[-\frac{\pi}{4}\left(\frac{\ep^{2/3}}{2\pi}\frac{(k_0\xi)^{4/3}}{\xi^{4/3}}\right) t_{-}^2 \right]} - \exp {\left[-\frac{\pi}{4}\left(\frac{\ep^{2/3}}{2\pi}\frac{(k\xi)^{4/3}}{\xi^{4/3}}\right) t_{-}^2 \right]}\right\}\\
 = & \,\, \frac{4\bar{w}^2}{5}(4\pi^2) C_k^2 \ep^{4/3}\xi^{13/3}\left(\frac{32\xi^{4/3}}{\ep^{2/3}t_-^2}\right) \exp {\left(-\frac{\pi}{4}\eta_{k}^{2} t_{-}^2 \right)}(k\xi)^{-17/3}\\
 &\times\left[\exp {\left(-\frac{\pi}{4}\eta_{k_0}^{2} t_{-}^2 \right)} - \exp {\left(-\frac{\pi}{4}\eta_{k}^{2} t_{-}^2 \right)}\right]
\end{split}
\end{equation*}
When $q \gg k$, approximation follows as $\tl{q}\approx q$, giving:
 
\begin{equation*}
\begin{split}
\Pi_B (k, t_1, t_2) = & \,\, 4\bar{w}^2  \int_{-1}^{1} d\mu  \left(1-\mu^2 \right)^2  \int_{k}^{\infty} \frac{q^2 dq}{(2\pi)^2} \left[P_v(q)\right]^2 \exp{\left[-\frac{\pi}{4}\times 2\eta_{q}^{2} t_{-}^2 \right]}\\
 = & \,\, 4\bar{w}^2  \int_{-1}^{1} d\mu  \left(1-\mu^2 \right)^2  \int_{k}^{\infty} \frac{q^2 dq}{(2\pi)^2} \left[4\pi C_q\ep^{2/3}q^{-11/3}\right]^2 \exp{\left[-\frac{\pi}{2}\eta_{q}^{2} t_{-}^2 \right]}\\
 = & \,\, 4\bar{w}^2 4\pi^2 \frac{16}{15}\ep^{4/3}\int_{k}^{\infty} dq C_q^2 q^2 q^{-22/3}\frac{\xi^{-16/3}}{\xi^{-16/3}}\exp{\left[-\frac{\pi}{2}\left(\frac{1}{2\pi}\ep^{2/3} q^{4/3}\times \frac{\xi^{4/3}}{\xi^{4/3}}\right) t_{-}^2 \right]}
\end{split}
\end{equation*}
 
\begin{equation*}
\begin{split}
\Rightarrow \tl{\Pi}_B (k\xi, t_1, t_2) =& \,\,\frac{16\bar{w}^2}{15}(4\pi)^2 \ep^{4/3}\xi^{16/3}\int_{k\xi}^{\infty} \frac{d(q\xi)}{\xi} C_q^2 (q\xi)^{-16/3}\exp{\left[-\frac{\ep^{2/3}}{4\xi^{4/3}}(q\xi)^{4/3} t_{-}^2 \right]}\\
=& \,\,\frac{16\bar{w}^2}{15}(4\pi)^2 C_k^2 \ep^{4/3}\xi^{13/3} \left[-\frac{3}{4}\left(\frac{\ep^{2/3}}{4\xi^{4/3}}t_{-}^2\right)^{13/4}\Gamma \left(-\frac{13}{4} , \frac{\ep^{2/3}t_{-}^2}{4\xi^{4/3}}(q\xi)^{4/3}\right)\right]_{k\xi}^{\infty}\\
=& \,\, \frac{4\bar{w}^2}{5}(4\pi^2) C_k^2 \ep^{4/3}\xi^{13/3} \left[4 \times \left(\frac{\ep^{2/3}}{4\xi^{4/3}}t_{-}^2\right)^{13/4}\Gamma \left(-\frac{13}{4} , \frac{\ep^{2/3}t_{-}^2}{4\xi^{4/3}}(k\xi)^{4/3}\right)\right]
\end{split}
\end{equation*}
 
As $z \equiv k\xi$ and 
$$\tl{\Pi} (z, t_1, t_2) = \tl{\Pi}_A (z, t_1, t_2) + \tl{\Pi}_B (z, t_1, t_2)$$
Therefore,
$$\tl{\Pi}=\frac{(4\pi\bar{w})^2}{5} C_k^2 \ep^{4/3}\xi^{13/3} \left\{\left(\frac{32\xi^{4/3}}{\ep^{2/3}t_-^2}\right)\frac{e^\alpha (e^{\alpha_0}-e^\alpha)}{(k\xi)^{17/3}}+ \left[\left(\frac{\ep^{2/3}}{4\xi^{4/3}}t_{-}^2\right)^{13/4}\Gamma \left(-\frac{13}{4} , \frac{\ep^{2/3}t_{-}^2}{4\xi^{4/3}}(k\xi)^{4/3}\right)\right]\right\}$$
where,
$$\alpha =  \exp {\left(-\frac{\pi}{4}\eta_{k}^{2} t_{-}^2 \right)} \text{  and  } \alpha_0 =  \exp {\left(-\frac{\pi}{4}\eta_{k_0}^{2} t_{-}^2 \right)}$$
Since:
$$ \eta_{k|_{k \rightarrow z}}^2 = \frac{\ep^{2/3}}{2\pi\xi^{4/3}} z^{4/3} \text{     and     } \ep = \frac{\kappa_0}{(3C_k)^{3/2}}\frac{\bar{U}^3}{\xi} \text{     where,     } \kappa_0 \equiv k_0\xi$$
The quantity outside the bracket becomes:
 
\begin{equation*}
\tl{\Pi}_0 = \frac{(4\pi\bar{w})^2}{5} C_k^2 \ep^{4/3}\xi^{13/3} = \frac{(4\pi\bar{w})^2}{5} C_k^2 \frac{\kappa_0^{4/3}}{9C_k^2}\bar{U}^4 \xi^{9/3}
\end{equation*}
$$\Rightarrow \tl{\Pi}_0 = \frac{(4\pi\bar{w})^2}{5} \kappa_0^{4/3}\bar{U}^4 \xi^3 $$
First term in the bracket becomes:
$$\tl{\Pi}_1 = \left(\frac{32\xi^{4/3}}{\ep^{2/3}t_-^2}\right)\exp {\left[-\frac{\pi}{4}\eta_{k}^{2} t_{-}^2 \right]}_{k\rightarrow z} \left[\exp {\left(-\frac{\pi}{4}\eta_{k_0}^{2} t_{-}^2 \right)} - \exp {\left(-\frac{\pi}{4}\eta_{k}^{2} t_{-}^2 \right)}\right]_{k\rightarrow z}(k\xi)^{-17/3}$$
Simplifying:
 
\begin{equation*}
\begin{split}
\exp {\left(-\frac{\pi}{4}\eta_{k_0}^{2} t_{-}^2 \right)} = & \,\, \exp {\left[-\frac{\pi}{4}\left(\frac{\ep^{2/3}}{2\pi}\frac{(k_0\xi)^{4/3}}{\xi^{4/3}}\right) t_{-}^2 \right]}\\
= & \,\, \exp {\left[-\frac{\pi}{4}\frac{\ep^{2/3}}{2\pi}\frac{(\kappa_0)^{4/3}}{\xi^{4/3}} t_{-}^2 \right]}  = \exp {\left[-\frac{\pi}{4}\frac{\ep^{2/3}}{2\pi}\frac{t_{-}^2 }{\xi^{4/3}} \kappa_0^{4/3} \right]}
\end{split}
\end{equation*}
 
\begin{equation*}
\begin{split}
\tl{\Pi}_1 = & \,\, \frac{32\xi^{4/3}}{z^{17/3}t_-^2}\left[\frac{(3C_k)^{3/2}}{\kappa_0}\frac{\xi}{\bar{U}^3}\right]^{2/3}\exp {\left(-\frac{\pi}{4}\frac{\ep^{2/3}t_{-}^2}{2\pi\xi^{4/3}} z^{4/3}\right)}\\
& \times \left[\exp {\left(-\frac{\pi}{4}\frac{\ep^{2/3}t_{-}^2}{2\pi\xi^{4/3}} \kappa_0^{4/3}\right)}-\exp {\left(-\frac{\pi}{4}\frac{\ep^{2/3}t_{-}^2}{2\pi\xi^{4/3}} z^{4/3}\right)}\right]\\
= & \,\, \frac{96 C_k}{\kappa_0^{2/3}}z^{-17/3}\frac{\xi^2}{t_-^2\bar{U}^2}\exp {\left[-\frac{\kappa_0^{2/3}}{24C_k}\left(\frac{t_{-}^2\bar{U}^2}{\xi^{2}}\right)z^{4/3}\right]}\\
& \times \left\{\exp {\left[-\frac{\kappa_0^{2/3}}{24C_k}\left(\frac{t_{-}^2\bar{U}^2}{\xi^{2}}\right)\right]} {\kappa_0^{4/3}} -  \exp {\left[-\frac{\kappa_0^{2/3}}{24C_k}\left(\frac{t_{-}^2\bar{U}^2}{\xi^{2}}\right)\right]} {z^{4/3}}\right\}
\end{split}
\end{equation*}
Let $\zeta \equiv t_-/\tau$ where, $\tau = \xi / \bar{U}$. this gives:
Defining:
$$\kappa = \frac{\kappa_0^{2/3}}{96 C_k}$$
$$\Rightarrow \tl{\Pi}_1 = \kappa^{-1}z^{-17/3}\zeta^{-2}\exp {\left(-4\kappa\zeta^2 z^{4/3}\right)}\left[\exp{\left(-4\kappa\zeta^2 \right)}{\kappa_0^{4/3}} - \exp{\left(-4\kappa\zeta^2 \right)}{z^{4/3}}\right]$$
Second term in the bracket becomes:
 
\begin{equation*}
\begin{split}
\tl{\Pi}_2 = & \,\, 4 \times \left(\frac{\ep^{2/3}t_{-}^2}{4\xi^{4/3}}\right)^{13/4}\Gamma \left(-\frac{13}{4} , \frac{\ep^{2/3}t_{-}^2}{4\xi^{4/3}}z^{4/3}\right)\\
 = & \,\, 4 \times\left\{\frac{\left[\frac{\kappa_0}{(3C_k)^{3/2}}\frac{\bar{U}^3}{\xi}\right]^{2/3}t_{-}^2}{4\xi^{4/3}}\right\}^{13/4}\Gamma \left(-\frac{13}{4} , \frac{\left[\frac{\kappa_0}{(3C_k)^{3/2}}\frac{\bar{U}^3}{\xi}\right]^{2/3}t_{-}^2}{4\xi^{4/3}}z^{4/3}\right)
\end{split}
\end{equation*}
 
Recalling $\zeta \equiv t_-/\tau$:
$$\Rightarrow \tl{\Pi}_2 = 4 \times\frac{\kappa_0^{13/6}\zeta^{13/2}}{(12C_k)^{13/4}}\quad \Gamma \left(-\frac{13}{4} , \frac{\kappa_0^{2/3}}{12C_k}\zeta^2 z^{4/3}\right)$$
Gathering all the terms together and replacing:
$$\kappa = \frac{\kappa_0^{2/3}}{96 C_k}$$
everywhere:
 
\begin{multline*}
\tl\Pi (k\xi, t_-/\tau) = \frac{4\bar{w}^2}{45}(4\pi^2) \kappa_0^{4/3}\bar{U}^4 \xi^3 \left\{\frac{z^{-17/3}}{\kappa\zeta^2}\frac{\left[\exp{\left(-4\kappa\zeta^2 {\kappa_0^{4/3}}\right)} - \exp{\left(-4\kappa\zeta^2 {z^{4/3}}\right)}\right]}{\exp {\left(4\kappa\zeta^2 z^{4/3}\right)}} \right. \\
\left. + 4 \times(8\kappa)^{\frac{13}{4}}\zeta^{\frac{13}{2}}\Gamma \left(-\frac{13}{4}, \frac{8\kappa z^{\frac{4}{3}}}{\zeta^{-2}}\right)\right\}
\end{multline*}
 
The final expression for $\tl\Pi (k,t_1,t_2) = \tl{\Pi}(z,\zeta) = \tl{\Pi}$ is:
\[
\boxed{\tl\Pi =  \frac{(4\pi\bar{w})^2}{45}\kappa_0^{4/3}\bar{U}^4 \xi^3 \left\{\frac{z^{-17/3}}{\kappa\zeta^2} \left[e^{-4\kappa\zeta^2 \left({\kappa_0^{4/3} + z^{4/3}}\right)} - e^{-8\kappa\zeta^2 {z^{4/3}}}\right] + \frac{4\zeta^{13/2}}{(8\kappa)^{-\frac{13}{4}}}\Gamma \left(-\frac{13}{4}, 8\kappa \zeta^2 z^{\frac{4}{3}}\right)\right\}}
\]
Since the Gamma function is defined as:
$$\Gamma (s, x) = \int_x^\infty dt \quad t^{s-1} e^{-t}$$
Approximating - as $x \rightarrow 0$:
$$\Gamma (s, x) \approx -\frac{1}{s} x^s $$
and as $x \rightarrow \infty$:
$$\Gamma (s, x) \simeq x^{s-1} e^{-x} $$
Hence, the second term of the $\tl{\Pi}$ equation can be written as:
$$\Gamma \left(-\frac{13}{4}, \frac{8\kappa z^{\frac{4}{3}}}{\zeta^{-2}}\right) = + \frac{4}{13}  \left[8\kappa z^{4/3}\zeta^2\right]^{-13/4}$$
$$\Rightarrow \tl{\Pi}_2 = 4(8\kappa\zeta^2)^{13/4} \times \frac{4}{13}  \left[8\kappa z^{4/3}\zeta^2\right]^{-13/4} = \frac{16}{13}z^{-13/3}$$
Therefore, the approximated  $\tl{\Pi}$ function employed in the sound shell model is:
\[
\boxed{\tl\Pi =  \frac{(4\pi\bar{w})^2}{45}\kappa_0^{4/3}\bar{U}^4 \xi^3 \left\{\frac{z^{-17/3}}{\kappa\zeta^2} \left[e^{-4\kappa\zeta^2 \left({\kappa_0^{4/3} + z^{4/3}}\right)} - e^{-8\kappa\zeta^2 {z^{4/3}}}\right] +  \frac{16}{13}z^{-13/3}\right\}}
\]

\section*{$\Pi$ contours}
\label{contours}
The contour maps for the unequal time correlators of $\Pi$ for the reference models and the new model are plotted as below.
 
It can be seen that $\Pi$ starts being constant for very low $z$ but diverges as $z$ increases in all three cases.
\begin{figure}
\centering
\includegraphics[width=7cm]{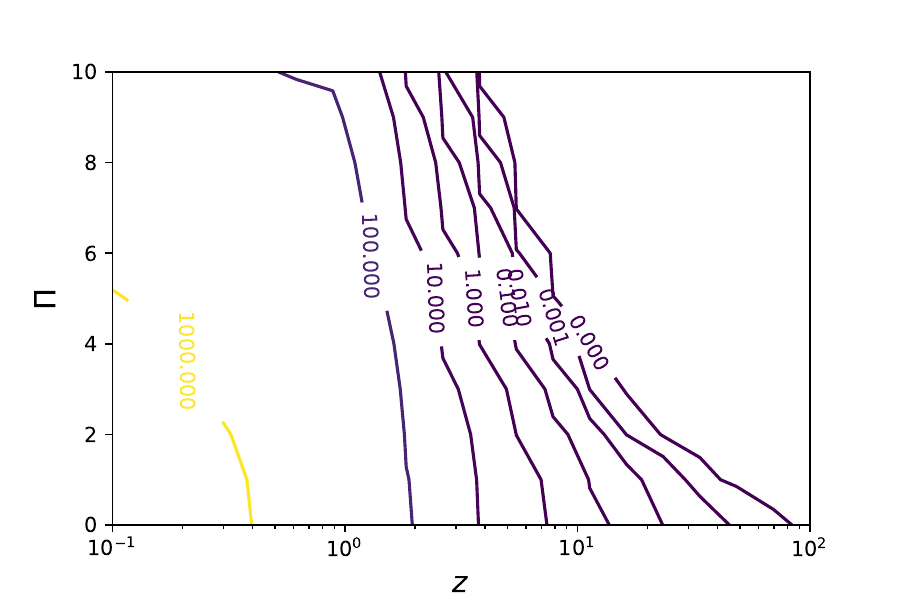}
\caption{\label{hkkc} Contour map of $\Pi$ with z and t for the first model}
\end{figure}
\begin{figure}
\centering
\includegraphics[width=7cm]{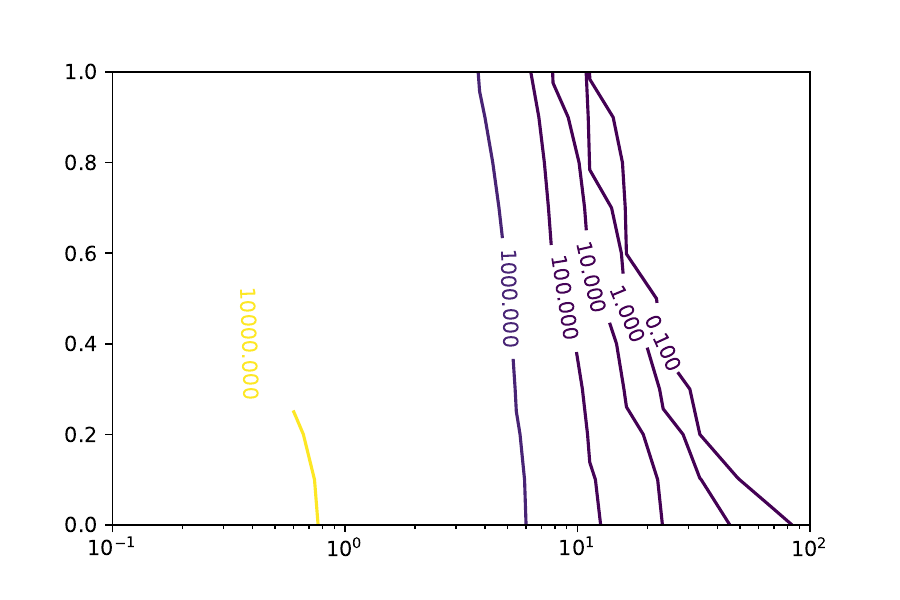}
\caption{\label{cdsc} Contour map of $\Pi$ with z and t for the second model}
\end{figure}
\begin{figure}
\centering
\includegraphics[width=7cm]{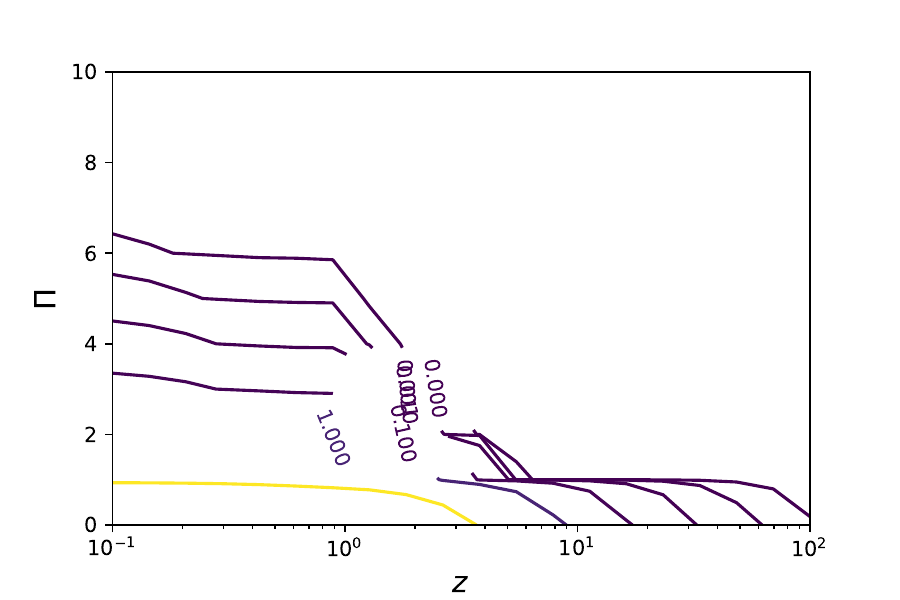}
\caption{\label{dtmc} Contour map of $\Pi$ with z and t for the new model}
\end{figure}

\chapter{Appendix: Python code}
\label{codea}

\section*{For finding the rate of spectral density, gravitational wave power spectrum and the amplitude vs frequency spectrum}
\begin{verbatim}
import numpy as np
from scipy import integrate
from scipy.integrate import dblquad
import matplotlib.pyplot as plt
import math

[REDACTED]
\end{verbatim}

\end{document}